\documentclass{aa}    
\usepackage{graphicx,url,twoopt,natbib}
\usepackage[varg]{txfonts}           
\usepackage{hyperref}                                   
\usepackage{natbib}								
\bibpunct{(}{)}{;}{a}{}{,}    
\usepackage{caption}							
\usepackage{multicol}							
\hypersetup{
  colorlinks=false,   
  urlcolor=blue,     
  linkcolor=red,     
}

\graphicspath{{Images/}}

\makeatletter
\renewcommand*\aa@pageof{, page \thepage{} of \pageref*{LastPage}}
\makeatother

\newcommand{\emptynote}[1]{{%
  \let\thempfn\relax
  \footnotetext[0]{#1}
}}

\begin{document}  

\titlerunning{Characterising Young M-dwarf Binaries}
\authorrunning{Calissendorff et al.}

\title{Characterising Young Visual M-dwarf Binaries with Near-Infrared Integral Field Spectra$^{*,\,**}$}

\author{Per Calissendorff$^{1}$ \and Markus Janson$^{1}$ \and Micka\"{e}l Bonnefoy$^{2}$
                                }

\institute{
        Department of Astronomy, Stockholm University, Stockholm, Sweden\\
        e-mail: {\bf per.calissendorff@astro.su.se}
\and
		Univ. Grenoble Alpes, CNRS, IPAG, 38000 Grenoble, France
}        

\date{Received 26 March 2020 / Accepted 21 July 2020}

\abstract{
We present the results from an integral field spectroscopy study of seven close visual binary pairs of young M-dwarf multiple systems. The target systems are part of the astrometric monitoring AstraLux program, surveying hundreds of M-dwarf systems for multiplicity and obtaining astrometric epochs for orbital constraints. Our new VLT/SINFONI data provides resolved spectral type classification in the $J, H$ and $K$ bands for seven of these low-mass M-dwarf binaries, which we determine by comparing them to empirical templates and examining the strength of water absorption in the $K$-band. The medium resolution $K$-band spectra also allows us to derive effective temperatures for the individual components. All targets in the survey display several signs of youth, and some have kinematics similar to young moving groups, or low surface gravities which we determine from measuring equivalent widths of gravity sensitive alkali lines in the $J$-band. Resolved photometry from our targets is also compared with isochrones from theoretical evolutionary models, further implying young ages. Dynamical masses will be provided from ongoing monitoring of these systems, which can be seen as emblematic binary benchmarks that may be used to calibrate evolutionary models for low-mass stars in the future.

}  

\maketitle



\section{Introduction} \label{sec:intro}
\emptynote{$^{*}$ Based on observations made with ESO telescopes at the Paranal Observatory under programme ID 0102.C-0186.}
\emptynote{$^{**}$ Reduced spectra can be downloaded at: \url{http://cdsweb.u-strasbg.fr/cgi-bin/qcat?J/A+A/}}

Binaries and higher-order hierarchical multiple systems of stars provide a plethora of useful information when studied. Population properties can provide insight to the different formation pathways and evolution, connecting higher-mass stars to their lower-mass counterparts, as well as to objects at the very low-mass bottom tail of the initial mass function, such as brown dwarfs. Keplerian motions in multiple systems reveal dynamical masses of the components, which together with their other physical features including age and brightness can be used to calibrate mass-luminosity relations \citep[e.g.][]{burgasser_not_2006, goodwin_relationship_2008, janson_improved_2007, kohler_orbits_2012, duchene_stellar_2013}. The more precisely these characteristics can be constrained, the better our physical interpretation becomes, forming some of our most fundamental understanding of stellar properties. Despite mid- to late-type M-dwarfs being the most numerous of stellar types in the solar neighbourhood, the fundamental physical characteristics such as age, mass, luminosity, radius and their relations are poorly constrained compared to intermediate mass and solar-type stars. Since multiplicity frequency decreases for later spectral types and lower primary mass, suitable benchmark multiple systems with detailed studies of orbital elements, and thereby dynamical masses, are lacking for low-mass M-dwarfs. As such systems are required for calibrating theoretical evolutionary models, significant effort have been placed in recent years to enhance characterisation of cool dwarfs, both for mid- to late M-type as well as for substellar \citep[e.g.][]{bouy_follow-up_2008, bonnefoy_young_2009, bonnefoy_library_2014, dupuy_studying_2010, schlieder_characterization_2014,  calissendorff_discrepancy_2017, calissendorff_improving_2018, calissendorff_spectral_2019}. Furthermore, with the recent abundant discoveries of planets with M-star hosts \citep[e.g.][]{dressing_occurrence_2013, gillon_temperate_2016}, accurate stellar physical properties are required in order to deriving reliable parameters for properties of the planetary systems, advancing the appeal to better characterise M-dwarfs \citep[e.g.][]{johnson_characterizing_2012, muirhead_characterizing_2012, muirhead_characterizing_2014, gaidos_trumpeting_2014, bowler_planets_2015}.

For young M dwarfs, typically below $\leq 100$ Myr, atmospheric and evolutionary models are particularly poorly constrained, since only a handful of young binaries have had their dynamical masses measured \citep[e.g.][]{close_dynamical_2005, bonnefoy_young_2009, montet_dynamical_2015, mizuki_orbital_2018, rodet_dynamical_2018}. However, one of the most difficult quantities to fully constrain is the age of a given system, which tends to dominate the uncertainty for most cases. Therefore, it becomes essential to identify systems where ages can be more easily assessed, such as associating them to kinematic groups with known ages, i.e. Young Moving Groups (YMGs), which are expected to be largely coeval \citep[e.g.][]{zuckerman_young_2004, torres_young_2008}. Due to the coevality assumption of these systems, they can be subjected to the more robust population age determination methods than what is possible for individual stars.

The goal of the AstraLux large M-dwarf survey is to improve constraints on multiplicity properties for M-dwarfs and identify young binaries that are suitable for orbital monitoring so that dynamical mass measurements can be carried out \citep{janson_astralux_2014, janson_orbital_2014}. By employing Lucky Imaging, the multiplicity survey discovered 761 nearby late K- and M-dwarfs \citep{bergfors_lucky_2010, janson_astralux_2012}, with an extension of the survey bringing 286 more later-type M-dwarfs \citep{janson_astralux_2014}. Recent studies of kinematics of YMGs and associations have revealed that a large fraction of the AstraLux targets are high probability members of these young associations. To better constrain the spectral types and youth of some of these M-dwarfs, we selected seven binary pairs or triple systems from the survey, five of which have strong indications of being members of YMGs, which were picked for integral field spectroscopic observations that are presented in this paper. These seven systems are all close visual binaries, which observed properties including luminosity, spectral type and surface gravity can be compared to theoretical evolutionary models of pre-main sequence stars, which in turn yield mass for individually resolved components and age estimates. As the systems are part of an ongoing astrometric monitoring survey, orbital constraints will be determined and thereby dynamical masses for these objects within a few years, and can be compared to the results presented here.

This paper is outlined as following after the introduction in Section~\ref{sec:intro}: Target description with observations and data reduction are described in Section~\ref{sec:obs_data}. In Section~\ref{sec:results} we present the methods applied for individual component spectral type characterisation and effective temperature measurements, as well as the procedure for measuring equivalent widths (EWs) of gravity sensitive features, which are then compared to old field stars and very young objects. In Section~\ref{sec:discussion} we discuss our results on individual systems, correlating different youth indicators and compare measured mass-luminosity relations. A summary and conclusions from this study is included in Section~\ref{sec:summary}.

\section{Observations \& Data Reduction} \label{sec:obs_data}
\subsection{Sample Selection \& Observations}
The observed sample consists of seven nearby M-dwarf binaries or higher hierarchical triple systems with a closely separated binary pair. These systems were chosen from the AstraLux M-dwarf survey target list and deemed as good candidates for follow-up observations with near-infrared spectroscopy and characterisation, similar to the target list composed by \citet{bergfors_characterisation_2016}. The targets are listed in Table~\ref{tab:obs} where names are otherwise abbreviated from the full Two Micron All-Sky Survey (2MASS) identifier to the form Jhhmm. All targets have been observed with Lucky Imaging by AstraLux over several epochs for probing common proper motion and are confirmed as physically bound, bona fide binaries. The binaries all have projected separations of $a \leq 9$ AU and are within distances $d \leq 60$ pc from the Sun. Furthermore, all targets are expected to be young based on their activity, YMG membership or low surface gravity measurements, most of them with expected ages ranging from 20 - 40 Myr, with the exception of J1036 which has been suggested to be part of the $\sim 400$ Myr Ursa Majoris moving group \citep{mamajek_discovery_2010, jones_ages_2015}. Four of our targets, J0459, J0611, J1036 and J2349 have had their parallax and proper motions measured by Gaia DR2 \citep{gaia_collaboration_gaia_2016, gaia_collaboration_gaia_2018}, but no data for J0111, J0332 or J1014 exist in the current release.

Based on the known kinematics of individual systems and using the Banyan $\Sigma$-tool \citep{gagne_banyan_2018} for estimating YMG membership probabilities, we find J0111 and J0332 to be part of the $\approx 24$ Myr old $\beta$pic moving group with $99.7\,\%$ and $97\,\%$ probabilities respectively. Previous estimates for the J0459 system with the BANYAN \texttt{II} \citep{gagne_banyan_2014} suggested it to be part of the $\approx$ 40 Myr Argus moving group, but the newer BANYAN $\Sigma$-tool finds with updated Gaia DR2 proper motions and parallax that the system is to be part of the field population with a $99.9\,\%$ probability. We find J0611 with a $95\,\%$ probability and J1014 with $92.5\,\%$ probability to both be part of the $\approx 45$ Myr old Carina moving group. However, the distance to the J1014 system is somewhat clouded in ambiguity. \cite{malo_bayesian_2013} suggests a statistical distance to the J1014 system of 69 pc based on its moving group membership, whereas \cite{riaz_identification_2006} finds a closer spectroscopic distance of 14 pc. Neither distance estimate has any affect on the BANYAN $\Sigma$ YMG membership probability. J1036 is a triple system with an equal brightness outer binary, and was previously assigned to the $\approx 400$ Myr Ursa Majoris moving group \citep{klutsch_reliable_2014}, but more recent kinematic measurements by Gaia DR2 and using the BANYAN $\Sigma$-tool suggests that the system is more likely to belong to the field, with a $99.9\,\%$ probability. Previous YMG membership probabilities from using the BANYAN \texttt{II} tool suggested the J2349 system to be a likely member of the $\beta$pic or Columba moving groups, with $94.8\,\%$ and $5.1\,\%$ probabilities respectively \citep{janson_orbital_2014}. Using the BANYAN $\Sigma$-tool on the other hand, we find with a $99\,\%$ that the system belongs to the field.

\begin{table*}[t]
\centering
\caption{Log of VLT/SINFONI observations.}
\begin{tabular}{lcccccc}
\hline \hline
2MASS ID & Alt. Name & Obs. Date & Exp. Time  & Exp. Time & Airmass  &Telluric STD \\
 & & & $J$ & $H + K$ & & SpT \\
\hline
2MASS J01112542+1526214 & GJ 3076 & 24-Dec-2018 & $5\times20$s & $5\times2$s & $\approx 1.4$ & B5\\
2MASS J03323578+2843554 & UCAC4 594–08941 & 27-Nov-2018 & $5\times30$s & $5\times30$s & $\approx 1.9$ & B2\\
2MASS J04595855-0333123 & UCAC4 433-008289 & 4-Nov-2018 & $5\times30$s & $5\times30$s &$\lesssim 1.2$ & B7 \\
2MASS J06112997-7213388 &  AL 442 & 28-Nov-2018 & $5\times30$s & $5\times30$s & $\lesssim 1.5$ & B4 \\
2MASS J10140807-7636327 & K2001c 27 & 24-Nov-2018& $5\times30$s & $5\times30$s & $\approx 1.7$ & B9\\
2MASS J10364483+1521394 & UCAC4 527-051290  & 26-Feb-2019 & $5\times30$s & $5\times10$s & $\approx 1.3$ & B4, B5\\
2MASS J23495365+2427493 & UCAC4 573-135909  & 9-Jun-2019 & $5\times30$s & $5\times30$s &$\approx 1.7$ & B2, B5\\
 \\
\hline
\label{tab:obs}
\end{tabular}
\end{table*}

Observations were carried out in service mode with the Adaptive Optics fed Spectrograph for INtegral Field Observations in the Near Infrared\footnote{SINFONI has since been decommissioned.} \citep[SINFONI;][]{eisenhauer_sinfoni_2003} at the Very Large Telescope Unit 4 (Yepun). The SINFONI instrument consists of the Spectrograph for Infrared Faint Field Imaging (SPIFFI) that is being fed by a modified version of the MultiApplication Curvature Adaptive Opticts system \citep[MACAO][]{bonnet_implementation_2003}. The observations were part of the {\bf ESO 0102.C-0186(A)} program and executed between November 2018 to June 2019. Exact dates for the observations of individual targets and their respective exposure times for each band are listed in Table~\ref{tab:obs}. All targets were observed with the $J (\lambda = 1.1 - 1.4\,\mu{\rm m})$ and $H\,+\,K\,(\lambda = 1.45 - 2.45\,\mu{\rm m})$ gratings, which have resolving powers of $R\,\approx\,2000$ and $R\,\approx\,1500$ respectively. The targets were sufficiently bright themselves and acted as natural guide stars. For most targets the Narrow Field Mode was employed, providing a spaxel scale of $12.5\,{\rm mas} \times 25\,{\rm mas}$ and a Field of View of $0.8\,{\rm arcsec}^2$, with the only exception being J0111, whose binary separation was too large and consequently observed with the larger $8\,{\rm arcsec}^2$ field of view and $125\,{\rm mas} \times 250\,{\rm mas}$ spaxel scale. Each source was observed with a small dithering pattern of five different positions in order to correct for bad pixels and to encompass all components, including the primary components for the triplet systems, J0332 and J1036. A skyframe was taken following the science observations, mimicking the exposure time of the science frame. The airmass ranged between 1.1 to 1.9 for separate targets, and each observation was accompanied by observations of a telluric standard star of spectral type B with similar airmass. Separate telluric standard stars were observed for J1036 for each band observed in, with a B4 spectral type for $J$-band and a B5 spectral type for $H+K$-band. Additionally, two telluric standard stars were observed together with J2349, due to the first standard star being a resolved visual binary itself, thus diluting the point spread function (PSF).


\subsection{Data reduction} \label{sec:red}
Raw data and associated calibration frames were compiled to construct $J$ and $H+K$ band datacubes using the ESO reduction pipeline \texttt{Esoreflex} \citep{freudling_automated_2013} for SINFONI, where all frames were combined to a final co-added datacube. An empirical polynomial law was applied to the data to correct for the target shift with wavelength induced by the atmospheric refraction. The quality of the data was inspected by eye, where we noted that for J2349 only the first frame was usable, and the component positions were either outside or partly outside the field of view in the remaining frames, causing a streak of low-signal pixels to move through the secondary component. Thus, only the datacube from the first frame for J2349 was used for further analysis.

All components in the resulting final datacubes were resolved with sufficient signal to noise that individual spectral elements could be extracted. Nevertheless, for these close binaries cross-contamination is an issue, which we addressed by applying the same procedure as presented in \cite{calissendorff_spectral_2019}, where a custom spectral extraction script is made to deblend the flux of the components for each wavelength slice within the datacube. The script is based on a grid-search in $x$ and $y$ positions (right ascension and declination) in the field of view, where a bright standard star is used as a PSF reference to find the optimal positions, and simultaneously scaled in brightness to match the respective component. We then subtract the PSF reference from one of the component positions and clean the image by removing the halo with a smoothed Gaussian profile fitted to the remaining binary component. The procedure is then repeated for the second component to minimise cross contamination. A model system is then produced, where the PSF reference is placed at the positions for the components and scaled to match their respective brightness, which is successively subtracted from the original frame to procure a residual image that can be compared to other positions and brightness scaling to find the optimal grid parameters. 

Since the PSF at the time of the observations of the targets were not necessarily the same as the PSF of the standard star observed just before or after, we considered a second approach as well, where we constructed a PSF reference by duplicating the brightest component in the target system, similar to what is done by \cite{bergfors_characterisation_2016}. We adopt the same naming of the two different approaches as \cite{bergfors_characterisation_2016}, calling the standard star reference as \emph{PSFstd} and the duplicated primary component reference as \emph{PSFdup}. For all considered cases, the \emph{PSFdup} reference achieved the smallest residuals from our PSF-matching algorithm for spectral extraction, and therefore considered more accurate. We thus performed all following calculations on the data sets where the \emph{PSFdup} reference was applied. Nevertheless, given the proximity to the secondary component for some systems, the \emph{PSFdup} reference may contain higher contamination during the deblending process described above.

Additionally, we separated the $H$ and $K$ band by considering only slices in the wavelength span of $\lambda = 1.45 - 1.80\,\rm{\rm m}$ for the $H$-band and $\lambda = 1.945 - 2.415\,\mu\rm{m}$ for the $K$-band. We also pruned the $J$-band to wavelengths $\lambda = 1.10 - 1.35\,\mu\rm{m}$. The choice of wavelength ranges was made in order to minimise inclusion of strong telluric water regions and to remove the edges of the pass bands where SINFONI performs poorly. 

The procedure is sensitive to how closely the binary components are separated, especially in the {\it PSFdup} scenario when one component is acting as the reference for the other. For certain cases, for instance as for J1036, the deblending process may be less accurate when one component is close to the edge of the detector and not the entire halo is within the field of view.

Circular aperture photometry was performed for both science targets and telluric standards to obtain spectra, where aperture sizes were kept the same for targets as for their respective standard stars to minimise differential flux losses. Typical aperture size was set to $\approx 10 - 20$ spaxels in radius, centred upon the photocentre of the binaries in order to obtain unresolved fluxes. The size of aperture radius was decided depending on the projected separation between the binary components for each individual system, with a greater aperture size for a larger separation. Thus, undesired noise may enter the aperture when the separation and thereby aperture was large, contributing to the uncertainty for our spectral analysis. The unresolved flux was thereafter scaled according to the brightness scaling parameter derived from the PSF spectral extraction tool described above. The telluric standard star spectra were then cleaned from strong absorption lines, including the Pa$\beta$ line in the $J$-band and the Br-series in the $H+K$ band. We first divided the standard star spectra by a transmission spectrum from the ESO sky model tool\emph{SkyCalc}, \footnote{\url{http://www.eso.org/observing/etc/bin/gen/form?INS.MODE=swspectr+INS.NAME=SKYCALC}} and divided the resulting standard star spectra by a blackbody function of the corresponding temperature for the standard star. The resulting spectral response is then a flat spectrum containing mainly stellar and instrumental features. We sequentially fit a Gaussian profile to each strong spectral feature, and interpolate the continuum from either side of the feature over it, starting from a full width half maximum length outward from its centre. The previous steps simplify the Gaussian fitting procedure, as the spectra are flat and free from strong telluric features at wavelengths near the Hydrogen lines of interest. The spectra are then again multiplied by the sky model transmission spectrum, resulting in a stellar absorption free spectral response which the observed binary spectra are thereafter divided by to obtain our final science spectra.

\section{Methods}\label{sec:results}
\subsection{Spectral Type}\label{sec:spt}
The continuum of near-infrared spectra is sensitive to temperature changes, and also possesses strong atomic and molecular line features, and is therefore useful for probing spectral types of cool stars such as those in our sample. We follow the spectral best-fit procedure described in \cite{cushing_atmospheric_2008} to derive continuum spectral types of our resolved binary components, comparing them to empirical template spectra of cool M-dwarfs from the NASA Infrared Telescope Facility (IRTF) Spectral Library \citep{cushing_infrared_2005, rayner_infrared_2009} and younger late M-types and L-types from the Montreal Spectral Library \citep{gagne_banyan_2015}. The empirical templates used for comparison are listed in Table~\ref{tab:emps}. The observed spectra are first resampled to match the resolution of the template, and the Goodness-of-fit value, $G$, is calculated accordingly to

\begin{equation}\label{eq:gof}
G = \sum_{i=1}^{n}\left(\frac{f_i - C F_{ i}}{\sigma_i} \right)^2,
\end{equation}

for each individual model, where $n$ is the number of data points, or wavelength slices in this case, with $i$ being the wavelength index. $f_i$ and $F_i$ are the observed and model flux respectively, with $\sigma_i$ being the uncertainty, and $C$ is a scaling factor calculated as

\begin{equation}
C = \frac{\sum f_i F_{i} / \sigma_i^2}{\sum F_{i}^2 / \sigma_i^2}.
\end{equation}

\begin{figure*}
\centering
\begin{multicols}{2}
\includegraphics[width=\linewidth]{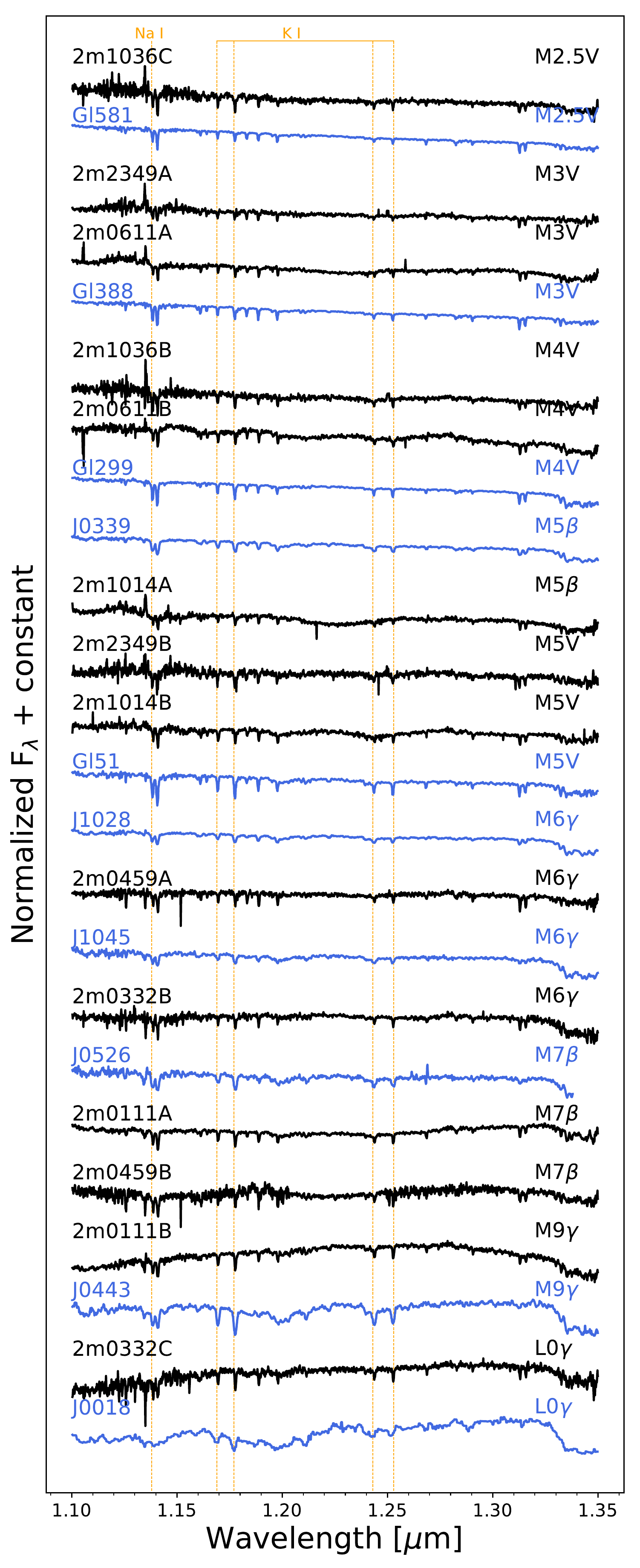}
\caption{$J$-band spectra of our observed binary targets in black, compared to IRTF template objects. The spectral types for our binaries are derived from the respective band continuum. Spectral features of special interest for our analysis are highlighted by the orange dashed lines.\label{fig:Jspec}}
\includegraphics[width=\linewidth]{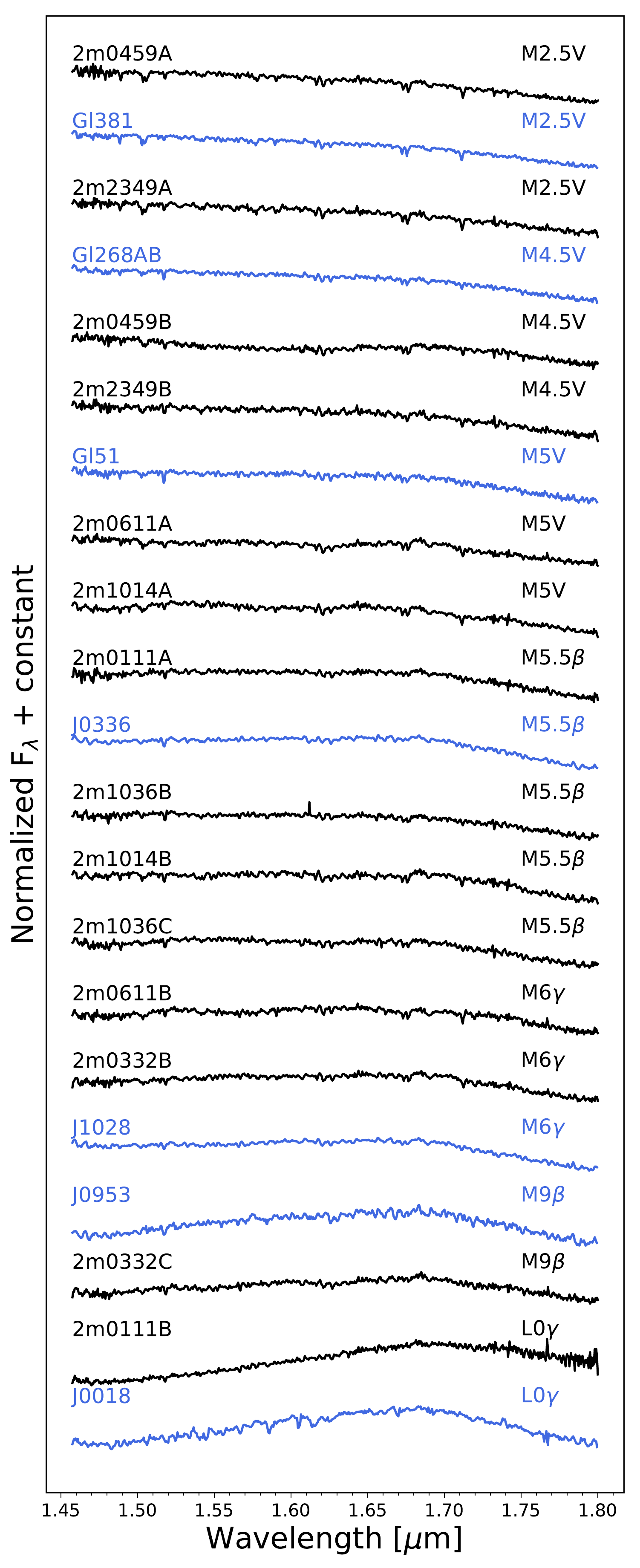}
\caption{Same as Figure~\ref{fig:Jspec} but for the $H$-band. Some spectra display irregularities around $1.73\,\mu$m which likely stems from the telluric removal procedure. \label{fig:Hspec}}
\end{multicols}
\end{figure*}

\begin{figure}[h!]
\includegraphics[width=\linewidth]{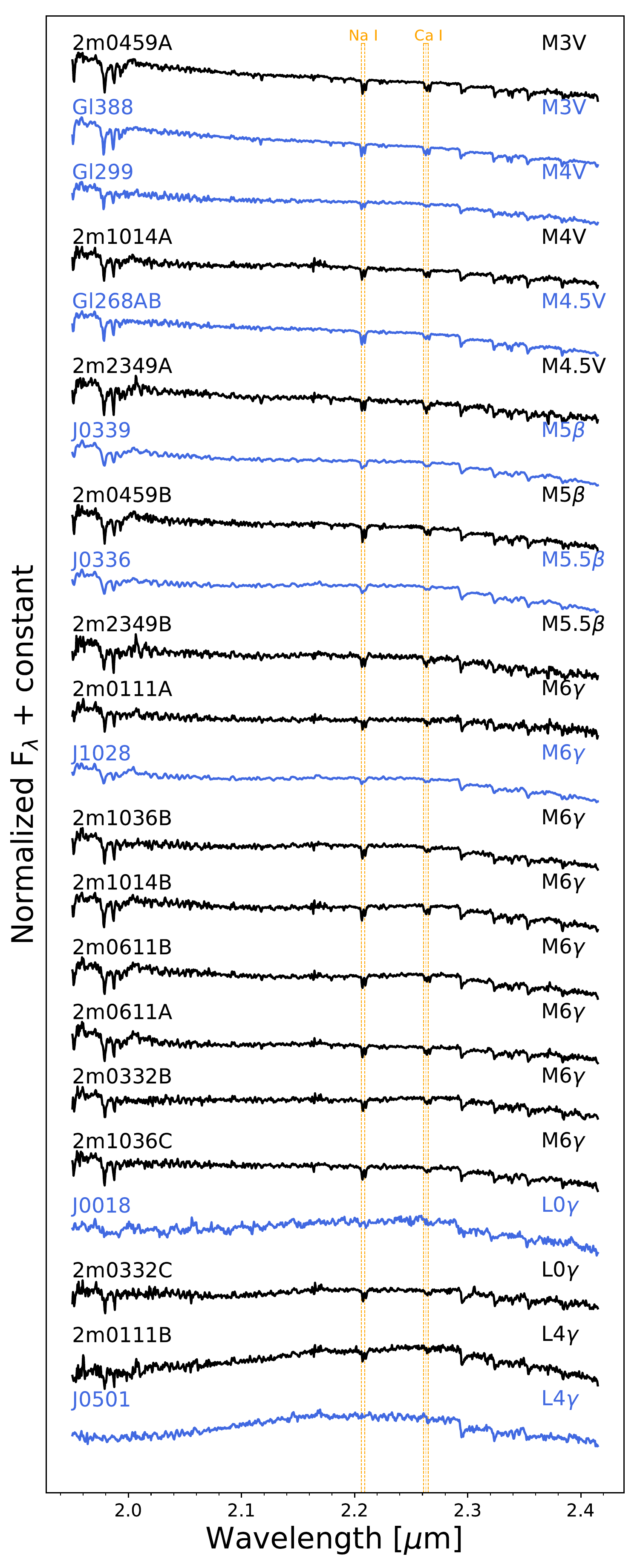}
\caption{Same as Figure~\ref{fig:Jspec} but for the $K$-band. The bump displayed by several of our spectra at $\approx 2.16\,\mu$m is likely due to a residual from the Br-$\gamma$ line in the telluric standards.\label{fig:Kspec}}
\end{figure}

For our continuum shape spectral analysis, we assign the template with the lowest calculated $G$-value as the best fit. We do this separately for all three bands considered, giving the resulting best-fit spectral type in Table~\ref{tab:spt}, and show all binary spectra together with the IRTF models for each respective band in Figures~\ref{fig:Jspec}, ~\ref{fig:Hspec} \&~\ref{fig:Kspec}, where the observed spectra have not yet been convolved to match their respective best-fit template. The plotted spectra were extracted using the \emph{PSFdup}-method described earlier, as it preserved the individual spectral shapes better. 

The IRTF templates used here are not listed as known members of any YMG, and are likely to be field stars and older than our target sample, therefore not necessarily reflecting the best fit to the spectral types of our sample. Nevertheless, the main difference between the IRTF models and our sample should be apparent in the line strength of gravity-sensitive features, with the IRTF templates having greater surface gravity values and thereby deeper lines. The benefit of comparing our target sample to the IRTF templates materialises from the similar resolving power of the SPeX spectrograph (R $\sim 2000$), wavelength coverage and high signal to noise ratio of the templates (SNR $\sim 100$), as well as consistency to previous work on AstraLux binaries \citep[e.g.][]{bergfors_characterisation_2016}. The Montreal Spectral Library however, provides late M and L spectral types with several youth indicators such as low surface gravity, specifying subtypes with low and intermediate surface gravities, $\gamma$ and $\beta$, that our target sample can be tested against.

We complemented the initial spectral estimate by adopting a more quantitative analysis of spectral types by \cite{rojas-ayala_metallicity_2012}, measuring the strength of the temperature dependent $K$-band water absorption as the H$_2$O $-$ K2 index, calculated from the median flux in each wavelength region by

\begin{equation}\label{eq:H2O}
H_2O - K2 = \frac{\left< F(2.070 - 2.090)\right> / \left< F(2.235 - 2.255)\right>}{\left< F(2.235 - 2.255)\right> / \left< F(2.360 - 2.380) \right>}.
\end{equation}

The equation for calculating the H$_2$O -- K2-index here is a modified version from \cite{covey_age_2010}, which here takes the Mg \texttt{I} and Ti \texttt{I} atomic features that affect bright spectral measurements into account. The $1-\sigma$ uncertainty of the H$_2$O -- K2-index was calculated from a Monte Carlo simulation with random Gaussian noise based on the signal to noise ratio of the individual spectrum. The simulation was repeated 10 000 times where the standard deviation from the measured H$_2$O -- K2-index was used as the error. The resulting indices are listed in Table~\ref{tab:spt} and used to calculate spectral types according to 
\begin{equation}\label{eq:H20_spt}
M_{\rm subtype} = 24.699 -23.788 ({\rm{H}_2{\rm O - K}2}),
\end{equation}
where the root mean square error for the spectral subtype is $\pm 0.624$. The calculated spectral types are listed in Table~\ref{tab:spt} together with the goodness-of-fit spectral types for each band. However, the spectral types derived from Equation~\ref{eq:H20_spt} may not reflect the true spectral types of the objects here, as the relation from \citep{rojas-ayala_metallicity_2012} is derived from a sample of field dwarfs, likely to be older than the our target sample. We also include previous spectral type designation from the literature, where \cite{janson_astralux_2012} derived spectral types using photometry in the $z'$ and $i'$ bands, following the procedure described in \cite{bergfors_lucky_2010}, which was developed by \cite{daemgen_discovery_2007} based on the relationship between magnitudes and spectral types presented by \cite{kraus_stellar_2007}.

\begin{table*}[t]
\centering
\caption{Near-infrared spectral types and effective temperatures for our binary pairs.}
\begin{tabular}{lccccccccc}
\hline \hline
Name & Literature$^{\dagger}$ & $J$-band & $H$-band & $K$-band & $K$-band & H2O-K2 & $K$-band  & H2O-K2  & H2O-K2 \\
 & SpT$(\pm 0.5)$ & SpT & SpT & SpT & SpT $(\pm 0.6)$ & $T_{\rm eff} [\rm{K}]$ & $\log g$ & $T_{\rm eff} [\rm{K}]$  & index \\
\hline
J0111A & M5.0 & M7$\beta$ & M$5.5\beta$ & M$6\gamma$ & M3.1 & $2900_{-200}^{+200}$ & 4.0 & $3564 \pm 29$ & $ 0.906 \pm 0.009 $ \\
J0111B & M6.0 &M$9\gamma$ & L$0\gamma$ & L$4\gamma$ & M9.6 & $2300_{-100}^{+200}$ & 3.5 & $2428 \pm 25$ & $ 0.635 \pm 0.007 $ \\
J0332B & M4.5 & M$6\gamma$ & M$6\gamma$ & M$6\gamma$ & M4.7 & $2800_{-200}^{+300}$ & 4.0 & $3456 \pm 30$ & $ 0.841 \pm 0.014$ \\
J0332C & M5.5  & L$0\gamma$ & M$9\beta$ & L$0\gamma$ & M5.5 & $2700_{-200}^{+300}$ & 3.5& $3371\pm 60$ & $ 0.808 \pm 0.018$ \\
J0459A & M4.0 & M$6\gamma$ & M2.5V & M3.0V & M1.7 & $3400_{-100}^{+100}$ & 4.5 &$3734 \pm 59$ & $0.966 \pm 0.011$ \\
J0459B & M5.5 & M$7\beta$ & M4.5V & M$5\beta$ & M4.5 & $3100_{-100}^{+200}$ & 4.5 & $3270 \pm 70$ & $0.849 \pm 0.016$  \\
J0611A & M4.0 & M3V & M5V & M$6\gamma$ & M2.9 & $3100_{-300}^{+200}$ & 4.5& $3657 \pm 34$ & $0.916 \pm 0.011$ \\
J0611B  & M5.0 & M4V & M$6\gamma$ & M$6\gamma$ & M5.2 & $3000_{-200}^{+200}$ & 4.5 & $3411 \pm 30$ & $0.820 \pm 0.011$ \\
J1014A & M4.0  & M$5\beta$ & M5V & M4V & M2.9 & $3100_{-200}^{+100}$ & 4.5 & $3660 \pm 42$ & $0.917 \pm 0.012$ \\
J1014B & M5.5  & M5V & M$5.5\beta$ & M$6\gamma$ & M5.2 & $2900_{-100}^{+200}$ & 4.5 & $3407 \pm 30$ & $0.818 \pm 0.011$ \\
J1036B & M5.0 & M4V & M$5.5\beta$ & M$6\gamma$ & M5.8 & $2900_{-100}^{+100}$ & 4.5 & $3051 \pm 41$ & $0.795 \pm 0.011$  \\
J1036C & M5.0 & M2.5V & M$5.5\beta$& M$6\gamma$& M4.3 & $2900_{-100}^{+200}$ & 4.5 & $3309 \pm 57$ & $0.858 \pm 0.013$ \\
J2349A & M3.5 & M3V & M2.5V & M4.5V & M4.1 & $3300_{-200}^{+100}$ & 4.5 & $3513 \pm 62$ & $0.866 \pm 0.023$ \\
J2349B & M4.5 & M5V & 4.5V & M$5.5\beta$ & M5.2 & $3100_{-200}^{+200}$ & 4.5 & $3412 \pm 102$ & $0.821 \pm 0.034$ \\
 \\
\hline
\label{tab:spt}
\end{tabular}
\par
{\scriptsize
{$^{\dagger}$} Spectral types derived by \cite{janson_astralux_2012} based on $(z' - i')$ photometry.
}
\end{table*}

\begin{table}[t]
\centering
{
\caption{Empirical templates employed for spectral fitting.}
\begin{tabular}{lcc}
\hline \hline
Name & Spectral type & Reference \\
\hline
HD 42581	 &	M1V & R09 \\
Gl 806	&	M2V & R09 \\
Gl 381	&	M2.5V & R09 \\
Gl 581	&	M2.5V &R09 \\
Gl 388	&	M3V & C05, R09 \\
Gl 273	&	M3.5V & R09 \\
G l213	&	M4V &  C05, R09 \\
Gl 299	&	M4V & R09 \\
Gl 268AB	&	M4.5V & R09 \\
Gl 51	 &	M5V &  C05, R09 \\
Gl 406	&	M6V &  C05, R09 \\
Gl 644C	&	M7V & C05, R09 \\
Gl 752B	&	M8V &  C05, R09 \\
LP 412-31	 & M8V & R09 \\
2MASS J07464256+2000321AB	 &	L0.5 &   C05 \\
2MASS J14392836+1929149	&	L1 &  C05 \\
2MASS J02081833+2542533	&	L1 &  C05 \\
Kelu-1AB	& L2 & C05 \\
2MASS J11463449+2230527AB	&	L3 & C05 \\
2MASS J15065441+1321060	&	L3 &  C05 \\
2MASS J00361617+1821104	&	L3.5 &  C05 \\
2MASS J22244381-0158521	&	L4.5 &  C05 \\
2MASS J15074769-1627386	&	L5 & C05 \\
2MASS J03390160-2434059	&	M5$\beta$ &  G15\\
2MASS J13582164-0046262	&	M5.5$\gamma$ &  G15\\
2MASS J03350208+2342356	&	M5.5$\beta$ &  G15\\
2MASS J10284580-2830374	&	M6$\gamma$ &  G15\\
2MASS J10455263-2819303	&	M6$\gamma$ &  G15\\
2MASS J08034469+0827000	&	M6$\beta$ &  G15\\
2MASS J20391314-1126531	&	M7$\beta$ &  G15\\
2MASS J05264316-1824315	&	M7$\beta$ &  G15\\
2MASS J03350208+2342356	&	M7.5$\beta$ &  G15\\
2MASS J08561384-1342242	&	M8$\gamma$ &  G15\\
2MASS J04433761+0002051&	M9$\gamma$ &  G15\\
2MASS J11064461-3715115	&	M9$\gamma$ &  G15\\
2MASS J09532126-1014205	&	M9$\beta$ &  G15\\
2MASS J00182834-6703130	&	L0$\gamma$ & G15\\
2MASS J23255604-0259508	&	L1$\gamma$ &  G15\\
2MASS J05361998-1920396	&	L2$\gamma$ &  G15\\
2MASS J04185879-4507413	&	L3$\gamma$ & G15\\
2MASS J05012406-0010452	&	L4$\gamma$ & G15\\
 \\
\hline
\label{tab:emps}
\end{tabular}
\par
{\scriptsize
{\bf References:}\\
 C05 = \citet{cushing_infrared_2005}, R09 = \citet{rayner_infrared_2009}, G15 =  \citet{gagne_banyan_2015}}
}
\end{table}

\subsection{Effective Temperature}\label{sec:Teff}
The near-infrared $K$-band is sensitive to temperature changes and an excellent probe for spectral classification. We estimate effective temperatures, $T_{\rm eff}$, solely from the $K$-band spectra, implementing two different approaches. Firstly, similar to the procedure described in Section~\ref{sec:spt} on how we determined spectral types from the continuum shape, we compare our observed resolved spectra to the BT-Settl CIFIST 2011-2015 theoretical evolutionary models \citep{baraffe_new_2015} by calculating the goodness-of-fit value using Equation~\ref{eq:gof} for a grid of different temperatures and surface gravities, with steps of $\Delta T_{\rm eff} = 100$ K and $\Delta \log g = 0.5$. The theoretical spectra are prior to the fit convolved to match the resolution of the observed SINFONI spectra, and we adopt the lowest goodness-of-fit value, $G$, as our best fit model for effective temperature, assigning those models with $G < \sqrt{2}\,G_{\rm best}$ as our uncertainty. The derived effective temperatures with uncertainty ranges are shown in Table~\ref{tab:spt}. The goodness-of-fit test takes no specific weight for certain areas of the spectra, and is a direct comparison between the overall shape of the observed and template spectra. We find that within the adopted uncertainty for $G$, models typically vary in surface gravities between $\log g = 3.5 - 5.0$, but when directly comparing only gravity sensitive lines, lower surface gravity values are favoured. \bigskip 

For the second approach we make use of the the H$_2$O -- K2-index calculated with Equation~\ref{eq:H2O}, which also works as a probe for temperature in the $K$-band. \cite{rojas-ayala_metallicity_2012} provide the results from using the H$_2$O -- K2-index as a temperature indicator for the 2010 version of the BT-Settl models, finding a monotonic function, insensitive of metallicity and surface gravity changes for stars $3000\,{\rm K} \leq T_{\rm eff} \leq 3800$ K \citep[see Table 4 in][]{rojas-ayala_metallicity_2012}. Below this boundary however, there is a clear difference between metal poor and metal rich stars, and although the cause of the metallicity behaviour at the lower bound is unknown, metallicity becomes an important factor for cool stars such as those in our sample that are expected to be near the lower temperature boundary. For our purposes we assume solar metallicity for the target sample, and similar to \cite{rojas-ayala_metallicity_2012} we derive a relation between effective temperature and the H$_2$O -- K2-index for the updated BT-Settl CIFIST 2011-2015 models, calculating the H$_2$O -- K2-index with Equation~\ref{eq:H2O} for each model with surface gravity of $\log g = 3.5,\,4.0\,\&\,4.5$, ranging effective temperatures $T_{\rm eff} = 2000 - 3800$ K, tabulated here in Table~\ref{tab:Teff}.

\begin{table}[t]
\centering
{
\caption{H2O-K2 index from BT-Settl CIFIST models.}
\begin{tabular}{lccc}
\hline \hline
$T_{\rm eff} [{\rm K}]$ & $\log g = 3.5 $ & $\log g = 4.0 $ & $\log g = 4.5 $ \\
\hline
2100 &	0.524 & 0.504 & 0.519\\
2200 &	0.547 & 0.531 & 0.533\\
2300 &	0.585 & 0.560 & 0.549\\
2400 &	0.627 & 0.600 & 0.578\\
2500 &	0.655 & 0.640 & 0.618\\
2600 &	0.680 & 0.678 & 0.656\\
2700 &	0.703 & 0.706 & 0.707\\
2800 &	0.719 & 0.730 & 0.732\\
2900 &	0.708 & 0.741 & 0.755\\
3000 &	0.714 & 0.754 & 0.780\\
3100 &	0.741 & 0.786 & 0.810\\
3200 &	0.762 & 0.801 & 0.833\\
3300 &	0.789 & 0.829 & 0.856\\
3400 &	0.815 & 0.860 & 0.879\\
3500 &	0.861 & 0.889 & 0.909\\
3600 &	0.895 & 0.916 & 0.942\\
3700 &	0.931 & 0.988 & 0.960\\
3800 &	0.946 & 1.006 & 0.977\\
 \\
\hline
\label{tab:Teff}
\end{tabular}
}
\end{table}

\subsection{Surface Gravity}
Age typically dominates the uncertainty for stellar parameters, and is one of the most difficult to determine accurately. Hence, a combination of several different youth indicators is typically required in order to qualitatively determine youth, as even though many individual signs may suggest a young age, they can not by themselves establish that the star is de facto young. Some of these indicators include chromospheric and coronal activity in the form of emission features such H$\alpha$, X-ray emission or flares, but also includes signs of accretion from photometric excess or forbidden O \texttt{I} emission lines. In the optical, absorption features of the Li \texttt{I} $6708\,\AA$ line can be used as a diagnostic for youth from lithium abundances or testing the lithium depletion boundary for low-mass stars. Another prominent method for estimating ages of stars is to adopt a \emph{guilt by association} approach, checking for kinematic properties that are consistent with known YMGs or associations. We discuss potential YMG affiliations further for individual systems in our sample in Section~\ref{sec:systems}.

Low surface gravity may also be used as a youth indicator. For instance, alkali lines in the near-infrared for a given spectral subtype are sensitive to gravity changes and will display a reduced strength compared to main sequence stars \citep[see e.g.][]{gorlova_gravity_2003, kirkpatrick_discovery_2006, allers_characterizing_2007}. \citet{mentuch_lithium_2008} surveyed lithium depletion for young stellar associations, including the $20 - 30$ Myr old Tucana-Horologium and $\beta$pic groups which several of our target objects are high probability members of. Using gravity sensitive spectral features, they report on surface gravities for most low-mass members to be around $\log g = 4.0 - 4.8 $, which is what we can expect from our target sample. Furthermore, \citet{david_age_2019} reports surface gravities of $\log g \approx 4.1$ for very young, $\leq 10$ Myr, low-mass binaries in Upper-Scorpius, and we can expect our target sample to have near but higher surface gravity values in comparison to Upper-Scorpius objects of similar mass.

Here we check our observed spectra for low surface gravity by measuring equivalent widths of the gravity sensitive Na \texttt{I} doublet at $1.138\,\mu$m and the K \texttt{I} lines at $1.169, 1.177, 1.243\,\&\, 1.253\,\mu$m in the $J$-band. The procedure is similar to what is described in \cite{bonnefoy_library_2014} and \cite{bergfors_characterisation_2016}, following the prescription of \cite{sembach_observations_1992}. Since M-dwarf spectra are greatly affected by molecular and broad band features in the continuum, we calculate a \emph{pseudo-continuum} using adjacent regions to the features of interest which are free of other atomic features. The EWs are calculated according to a Riemann sum expression of the integrated area over the wavelength region of interest as 
\begin{equation}\label{eq:EW}
EW_{\lambda} \simeq \sum_{i=0}^n \left[ 1 - \frac{{\rm F}(\lambda_i)}{{\rm F_c}(\lambda_i)} \right] \Delta \lambda_i,
\end{equation}
where ${\rm F}(\lambda_i)$ and ${\rm F_c}(\lambda_i)$ are the respective line flux and pseudo-continuum flux of the wavelength interval $\Delta \lambda_i$. 

The uncertainties for the EW measurements are obtained via a Monte Carlo approach where a Gaussian noise based on the signal-to-noise of the spectra is added to the observed spectra. The EW of each feature is then calculated using Equation~\ref{eq:EW} and the procedure is repeated $1 000$ times, where the standard deviation from the EW of the observed spectra is adopted as the uncertainty. The EWs and their uncertainties are listed in Table~\ref{tab:gravity}.

\begin{figure}
\includegraphics[width=\linewidth]{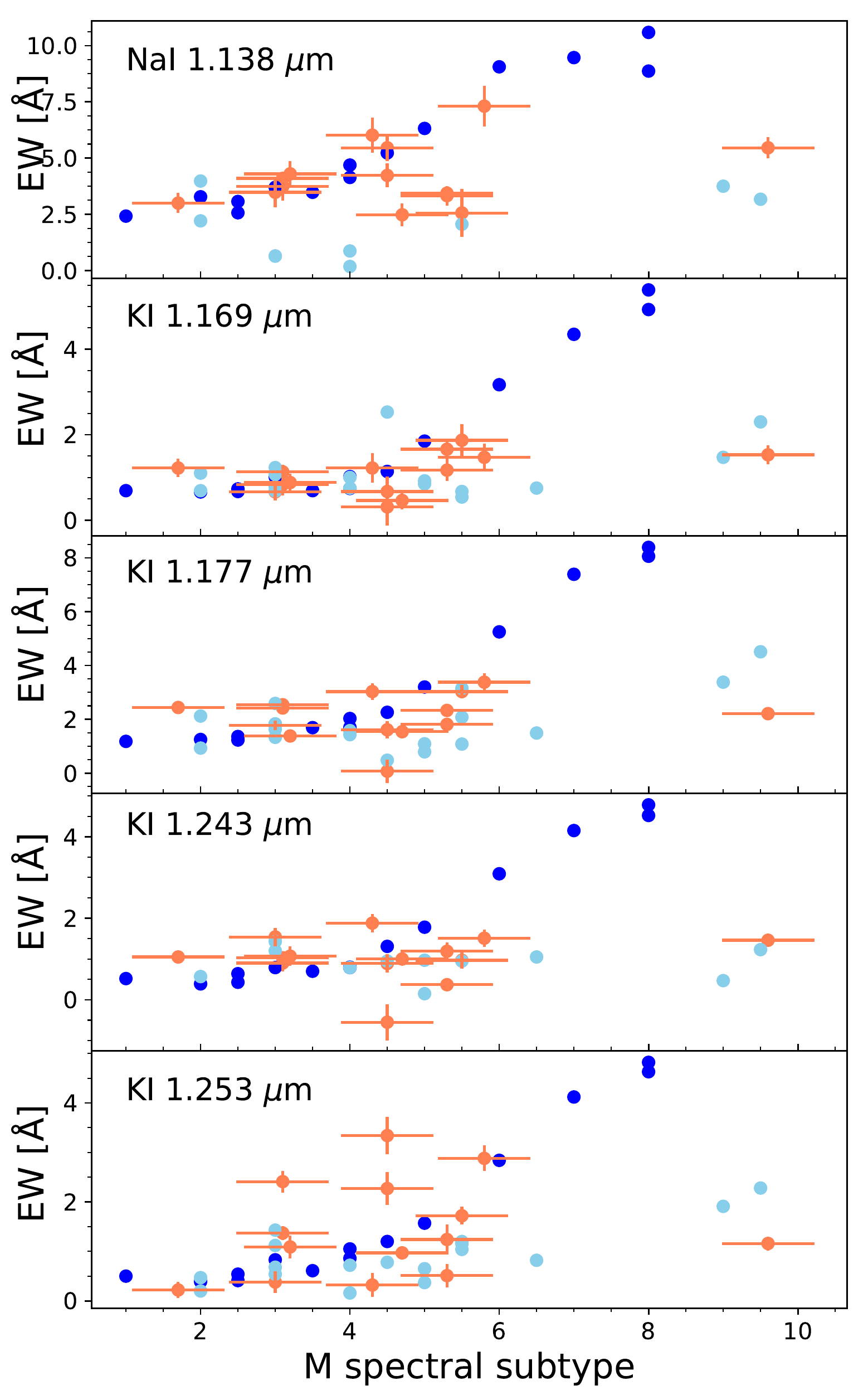}
\caption{Equivalent widths of surface gravity sensitive features, listed in Table~\ref{tab:gravity}. Our measured EWs from SINFONI spectra are depicted by the orange circles, and compared to measured EWs from older field stars in the IRTF sample, potrayed by the dark blue circles, and young $<10$ Myr stars from \cite{manara_x-shooter_2013}, represented by the light blue circles. The spectral types for our target sample are derived from the H$_2$O--K2-index in Table~\ref{tab:spt}.}
\label{fig:EWs}
\end{figure}

Equivalent widths are calculated using Equation~\ref{eq:EW}, where pseudo-continuum regions are fitted using Legendre polynomials of third order locally around the line of interest, which are then interpolated to the feature wavelength. The reference wavelengths used for the line feature and pseudo-continuum regions are shown in Table~\ref{tab:EWref}, where the pseudo-continuum regions are calculated using wavelength regions between $\lambda_1 - \lambda_2$ and $\lambda_3 - \lambda_4$, and the line feature from the region between $\lambda_2 - \lambda_3$. We list the measured EWs in Table~\ref{tab:gravity}, which are also plotted against spectral type in Figure~\ref{fig:EWs} together with EWs for IRTF SpeX library of field dwarfs and a sample of young stars $< 10$ Myr from \cite{manara_x-shooter_2013} for comparison. We find that objects in our sample with estimated spectral types of $\sim$M5 or later have systematically lower line strength, and thereby also lower surface gravity, compared to the older field dwarf IRTF sample. Conversely, our sample has stronger linestenght than the very young stars from \cite{manara_x-shooter_2013}, which is consistent with the estimated ages for our target sample, being part of YMGs and having ages intermediate to that of the old field and the very young populations. However, given the difficulties with the deblending procedure described in Section~\ref{sec:red}, it is possible that the lines are diluted by residual companion contamination and the uncertainties are underestimated.

We further test the validity of the derived surface gravities for our objects by comparing the observed spectra to synthetic spectra from the BT-Settl CIFIST 2011-2015 theoretical models, varying the surface gravities of the models at fixed metallicity and for specific temperatures. Best-fit calculations are performed using Equation~\ref{eq:gof} for constrained wavelength areas around the gravity sensitive features in the $J$-band shown in Table~\ref{tab:EWref}. The synthetic spectra indicate that lower temperature objects, $\lesssim 2800$ K, are more sensitive to changes in surface gravity, and we can test whether the observed SINFONI spectra are consistent with being young objects. At higher temperatures the sensitivity is more modest and the best-fit procedure develops into a degenerate issue. A discussion from the results of this test is presented Section~\ref{sec:discussion} for each individual system. 

\begin{table}[t]
\centering
{
\caption{Reference wavelengths for Na \texttt{I} and K \texttt{I} EWs.}
\begin{tabular}{lcccc}
\hline \hline
Feature & $\lambda_1$ &$\lambda_2$  & $\lambda_3$ & $\lambda_4$    \\
 & [$\mu$m] & [$\mu$m] & [$\mu$m] &[$\mu$m] \\
\hline
Na \texttt{I} $- 1.138 \mu$m & 1.1270 & 1.1360 & 1.1420 & 1.1580  \\
K \texttt{I} $- 1.169 \mu$m & 1.1560 & 1.1670 & 1.1710 & 1.1760 \\
K \texttt{I} $- 1.177 \mu$m & 1.1710 & 1.1750 & 1.805 & 1.1820 \\
K \texttt{I} $- 1.243 \mu$m & 1.2390 & 1.2415 & 1.2455 & 1.2490 \\
K \texttt{I} $- 1.253 \mu$m & 1.2455 & 1.2500 & 1.2550 & 1.2670\\
\hline
\label{tab:EWref}
\end{tabular}
}
\end{table}

\begin{table*}[t]
\centering
{
\caption{Equivalent widths of gravity sensitive features in the $J$-band.}
\begin{tabular}{lccccc}
\hline \hline
Name & Na \texttt{I} - $1.138\,\mu$m & K \texttt{I} - $1.169\,\mu$m & K \texttt{I} - $1.177\,\mu$m & K \texttt{I} - $1.243\,\mu$m & K \texttt{I} - $1.253\,\mu$m \\
 & [\AA] & [\AA] & [\AA] & [\AA] & [\AA] \\
\hline
J0111A &$4.10\pm0.17$&$1.13\pm0.13$&$2.54 \pm0.09$&$1.03 \pm 0.10$&$1.37\pm0.11$\\
J0111B &$5.46\pm0.41$&$1.53\pm0.19$&$2.21\pm0.11$&$1.46\pm 0.13$&$1.16\pm0.11$\\
J0332B &$2.48\pm 0.45$&$0.46\pm0.17$&$1.54\pm0.19$&$1.00\pm0.10$&$0.97\pm0.11$\\
J0332C &$2.56\pm1.01$&$1.87\pm0.35$&$3.03\pm0.20$&$0.97\pm0.17$&$1.72\pm0.15$\\
J0459A &$3.00\pm0.38$&$1.22\pm0.18$&$2.44\pm0.12$&$1.05\pm0.09$&$0.22\pm013$ \\
J0459B &$4.23\pm0.47$&$0.31\pm0.40$&$0.07\pm0.38$&$0.89\pm0.19$&$2.27\pm0.30$\\
J0611A &$4.30\pm0.50$&$0.88\pm0.17$&$1.38\pm0.14$&$1.07\pm0.20$&$1.09\pm0.20$\\
J0611B &$3.44\pm0.26$&$1.17\pm0.22$&$1.81\pm0.20$&$0.37\pm0.11$&$0.51\pm0.21$\\
J1014A &$3.48\pm0.60$&$0.66\pm0.17$&$1.77\pm0.12$&$1.54\pm0.19$&$0.39\pm0.19$\\
J1014B &$3.33\pm0.37$&$1.66\pm0.22$&$2.33\pm0.06$&$1.19\pm0.18$&$1.24\pm0.27$\\
J1036B &$7.31\pm0.83$&$1.47\pm0.29$&$3.38\pm0.28$&$1.51\pm0.18$&$2.88\pm0.23$\\
J1036C &$6.02\pm0.72$&$1.22\pm0.31$&$3.03\pm0.26$&$1.88\pm0.19$&$0.32\pm0.21$\\
J2349A &$3.74\pm0.57$&$0.83\pm0.22$&$2.42\pm0.14$&$0.90\pm0.17$&$2.41\pm0.19$\\
J2349B &$5.46\pm0.53$&$0.67\pm0.32$&$1.61\pm0.26$&$-0.55\pm0.41$&$3.34\pm0.35$ \\
 \\
\hline
\label{tab:gravity}
\end{tabular}
}
\end{table*}

\subsection{Astrometry \& Photometry}\label{sec:phot}
From the spectral extraction procedure described in Section~\ref{sec:red} we also obtained astrometric positions and resolved photometry for our binary components. To calibrate the photometric measurements of our targets we compare the unresolved 2MASS \citep{skrutskie_two_2006} magnitudes of our targets to magnitudes of the standard star observed together with the respective target. The magnitudes are sequentially scaled to the brightness difference of the components to obtain resolved photometry. Since the photometry is performed in a sub-set of wavelengths not entirely covering the full width of the 2MASS photometric measurements, albeit based on unresolved 2MASS data, we consider our astrometric and photometric measurements to be in pass bands of $J'\,(1.10 \leq \lambda_{J'} \leq1.35\,\mu\rm{m})$, $H'\,(1.45 \leq \lambda_{H'} \leq 1.80\,\mu\rm{m})$ and $K'\,(1.945 \leq \lambda_{K'} \leq 2.415\,\mu\rm{m})$. The difference in magnitudes for each binary pair is tabulated in Table~\ref{tab:phot} where the error is quadratically added from the 2MASS magnitude errors and the photon noise from the spectral extraction. Estimated magnitudes for all resolved components in the triple systems J0332 and J1036 are shown in Table~\ref{tab:triplet}.

We measure astrometric positions by collapsing the datacubes in the specific band of interest, and perform the same positional grid-search as we did for individual wavelength slices for the spectral extraction with the \emph{PSFdup} reference. The reason for collapsing the datacube in this context is to increase the signal-to-noise for more precise positions. We do this for each band individually, where the mean of the separation and positional angle for all three bands are shown in Table~\ref{tab:phot}, with the standard deviation from the mean as the uncertainty.

\begin{table*}[t]
\centering
{
\caption{Astrometric and photometric measurements for binary pairs.}
\begin{tabular}{lccccc}
\hline \hline
Name & sep & PA$^{\dagger}$ & $\Delta J'$ & $\Delta H'$ & $\Delta K'$\\
 & [mas] & [$^{\circ}$] & [mag] & [mag] & [mag] \\
\hline
J0111AB & $ 1040.9 \pm 13$ & $9.5 \pm 1.4$ & $0.98 \pm 0.03$& $1.30 \pm 0.05$ & $0.63 \pm 0.05$  \\
J0332BC  &$86.3\pm2.4$ &$19.6\pm1.4$& $0.87 \pm 0.02$& $0.74\pm 0.02$ & $0.65 \pm0.02$  \\
J0459AB & $142.6 \pm 2.0$ & $53.8 \pm 1.2$ & $1.19\pm0.02$&$1.50\pm0.03$ &$1.21\pm0.03$   \\
J0611AB   & $177.6\pm3.2$&$30.2\pm0.4$&$0.37\pm0.03$ &$0.31\pm0.03$ &$0.26\pm0.03$\\
J1014AB   &$283.9\pm0.1$ &$7.4\pm0.1$& $-0.05\pm0.03$&$0.05\pm0.03$ &$-0.02\pm0.02$   \\
J1036BC   &$99.8\pm2.9$ &$8.6\pm1.7$ &$0.02\pm0.02$ &$0.04\pm0.03$ &$-0.00\pm0.03$   \\
J2349AB &$208.1\pm1.1$ &$37.2\pm0.1$ &$1.13\pm0.03$ &$1.07\pm0.03$ & $1.07\pm0.02$  \\

 \\
\hline
\label{tab:phot}
\end{tabular}
\par{\scriptsize
$^{\dagger}$ Positional Angles have not been corrected for True North.
}
}
\end{table*}

\begin{table}[t]
\centering
\captionsetup{justification=centering}
\caption{Resolved photometry for triple systems.}
\begin{tabular}{lccc}
\hline \hline
Component &  $J'$ & $H'$ & $K'$\\
  & [mag] & [mag] & [mag] \\
\hline
J0332A & 9.83  & 9.45 & 9.16  \\
J0332B & 10.66 & 10.02 & 9.77  \\
J0332C & 11.53 & 10.76 &  10.42 \\
J1036A & 9.18 & 8.90 & 8.57  \\
J1036B & 10.71 & 9.70 & 9.48  \\
J1036C & 10.73 & 9.74 & 9.48 \\

 \\
\hline
\label{tab:triplet}
\end{tabular}
{\scriptsize \par
Magnitudes are scaled from their relative brightness and unresolved 2MASS photometry. Errors are on the order of $\pm 0.02$ mag.
}
\end{table}

\section{Results \& Discussion}\label{sec:discussion}
\subsection{Individual Systems}\label{sec:systems}
\cite{bergfors_lucky_2010} estimated that the AstraLux sample consists of stars younger than $\leq 1$ Gyr, with most being even younger than $\approx 600$ Myr based on their selection from the catalogue of stars with strong coronal emission and low tangential velocity by \cite{riaz_identification_2006}. More recently, large surveys of space velocities and kinematics have revealed that some of these AstraLux targets are high-probability members of YMGs and associations \citep[e.g.][]{malo_bayesian_2013, gagne_banyan_2014, gagne_banyan_2014-1}. Below we provide for each individual system in our sample a discussion on the findings in our survey as well as additional youth-indicators such as YMG membership.

\noindent {\bf 2MASS J01112542+1526214:} The system has the largest separation of the binary pairs in our sample, over 1''. We find from the spectral type and spectral EW analysis that the young age is consistent when compared to both the older field population as well as the very young sample. J0111B is the coolest object in our survey, and has an observed spectrum that our best-fit calculations find to be consistent with an L-type template. When we test synthetic spectra for spectral line feature changes with respect to change in surface gravity, we find a best-fit for low surface gravity, $\log g = 3.5$ for a low temperature of $T_{\rm eff} = 2300$ K for the binary component J0111B. We further find that although the strength of the spectral line features are compatible for higher temperatures models, the continuum fit is only consistent with low temperature models, which supports the notion of the object being young, and explains the difference in line strength compared to the IRTF models seen in Figure~\ref{fig:Jspec}.

\noindent {\bf 2MASS J03323578+2843554:} A triple system where the outer binary BC components are likely of later spectral types, $\sim$M6 and L0, according to both our continuum and spectral feature analysis. Both component B and C in the outer binary of J0332 are compatible with low surface gravity, and our best-fit test yields $\log g = 3.5$ for the pair, which also has relatively low effective temperatures of $T_{\rm eff} = 2800$ K and 2700 K respectively, consistent with a young age for the system. We find from the H2O-K2 index a discrepancy in the effective temperature measurement, obtaining higher values of $\approx 3400$ K. However, for these higher temperatures the continuum of the observed spectra deviates heavily from the synthetic models, and is more consistent with the lower temperature models with respect to both continuum and spectral line features.

\noindent {\bf 2MASS J04595855-0333123:} The system consists of one brighter $\sim$M2.5 primary with a $\sim$M4.5 secondary component. The $J$-band spectrum fit for the system shows a discrepancy compared to the other spectral type estimates. We find no obvious reason to why we obtain this discrepancy, but note that we obtain the worst fit for each component in this band, that perhaps a suitable template, for example a young and early M-type, was lacking. Our best-fit when comparing to theoretical model spectra suggest the primary and secondary components to have effective temperatures of $T_{\rm eff} = 3400$ and 3100 K respectively. We obtain a best-fit for models with surface gravity of $\log g = 4.5$, but at this temperature range lower surface gravities are also consistent with the shape of the spectra, and the gravity sensitive line features in the $J$-band are more consistent with a lower surface gravity of $\log g \approx 4.0$.

\noindent {\bf 2MASS J06112997-7213388:} We find for the continuum fit spectral types of M3-6 and M4-6 for the primary and secondary respectively, suggesting that our continuum fit was poorly constrained for this system. The fitted models within the adopted uncertainty of $\sqrt{2}$ of the lowest goodness-of-fit value range a mix of older IRTF and younger sub-spectral types, suggesting that early M-types younger than the templates probed here, $\leq M5\gamma/\beta$, could potentially be a better match for the system. The H$_2$O--K2-index analysis on the other hand suggests the primary to be of M$2.9\pm0.6$ spectral type, and M$5.2\pm0.6$ for the secondary. We find from our theoretical model best-fit that the components have effective temperatures of about 3100 K and 3000 K respectively, with the overall best goodness-of-fit for surface gravity of $\log g = 4.5$. The H2O-K2 index calculations suggest higher temperatures however, with effective temperatures of $T_{\rm eff} \approx 3600$ and 3400 for the individual components. The $J$-band individual spectral features show a better fit for slightly lower surface gravities, $\log g \approx 4.0 - 4.5$, consistent with the system being of young age.

\noindent {\bf 2MASS J10140807-7636327:} A well-separated binary with components of similar brightness. We obtain individual spectral types of $\approx$M5 and $\approx$M5.5 for the primary and secondary respectively. We find the best-fit effective temperatures for the components to be $T_{\rm eff} = 3100$ \& 2900 K respectively, with a best-fit surface gravity of $\log g = 4.5$. The $J$-band spectral feature analysis  however suggests lower surface gravities, and we obtain better fits for $\log g \lesssim 4.0$. However, we find better goodness-of-fit values for the individual spectral features in the $J$-band when applying synthetic models with higher $T_{\rm eff}$ values compared our $K$-band analysis, which temperatures are also more consistent with the values derived from the H2O-K2 index of $T_{\rm eff} \approx 3600$ and 3400 respectively.

\noindent {\bf 2MASS J10364483+1521394:} We estimate the spectral types for the outer binary to be between M2.5 - 6.0, with a favour towards later types from the continuum fit. Only the $J$-band spectra yield good fits for early M-types, which observations were of quite poor quality with low signal-to-noise ratio for the outer binary pair. We redo our spectral template fit for the outer binary of J1036 when it is unresolved, finding a best-fit for a M4V type. This may suggest that the signal to noise in the $J$-band was too low for our astrometric procedure and deblending process to provide accurate brightness scalings of the two components in the outer binary. The $J$-band spectrum was corrected from telluric contribution using a different standard star compared to the $H+K$ band, and the only system in our sample to use different standard telluric stars for different bands. \cite{calissendorff_discrepancy_2017} placed orbital constraints for the outer binary and estimated individual dynamical masses to be $\approx 0.24\,M_{\odot}$ for each component. The theoretical model best-fit procedure suggest an effective temperature of $T_{\rm eff} = 2900$ K for each component, and surface gravity of $\log g = 4.5$ that is consistent with the individual spectral line analysis. The EW analysis also show the strongest lines for this system, further supporting an older age estimate compared to the rest of our sample.

\noindent {\bf 2MASS J23495365+2427493:} We find from our continuum spectral fit that the primary is of M$3.5 \pm1.0$ spectral type and the secondary has spectral type M$5.0\pm0.5$, which are somewhat earlier than the estimates we obtain from the H$_2$O--K2-index analysis for each component. We estimate the effective temperature of the components as $T_{\rm eff} = 3300$ \& 3000 K respectively, with surface gravity of $\log g = 4.5$ . Similar surface gravities are found for the individual spectral line features, with a better fit tendency towards higher surface gravities $\approx 4.5.$. Both the empirical spectral template fit and synthetic theoretical models seem to suggest that the system could be somewhat older than the rest of our sample, and belong to the field rather than any known YMG. Nevertheless, the observations of 2M2349 have the highest uncertainty due to only one out of five frames being used in our analysis.

\subsection{Hertzsprung-Russel Diagram}
\begin{figure}[h!]
\centering
\includegraphics[width=\linewidth]{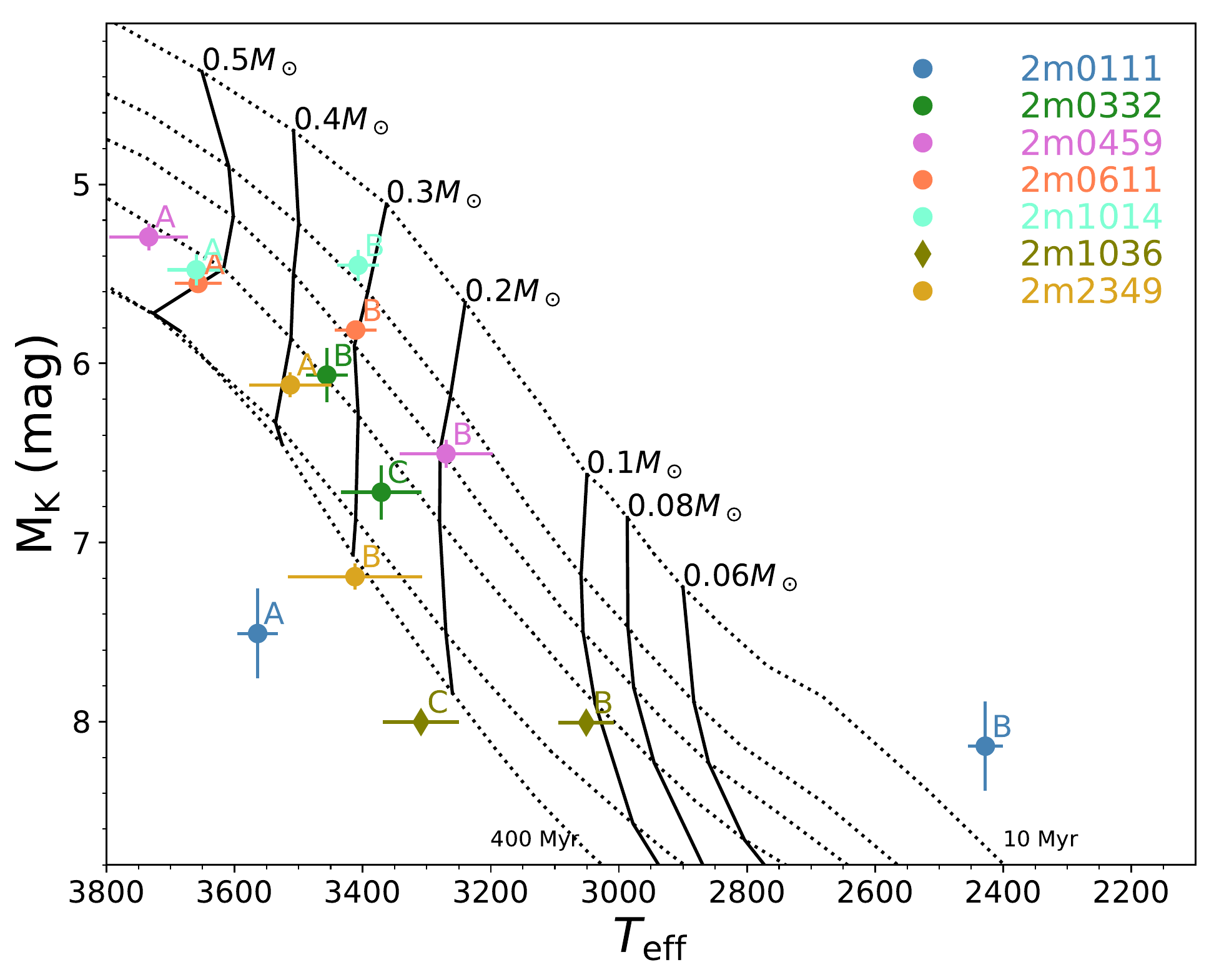}
\includegraphics[width=\linewidth]{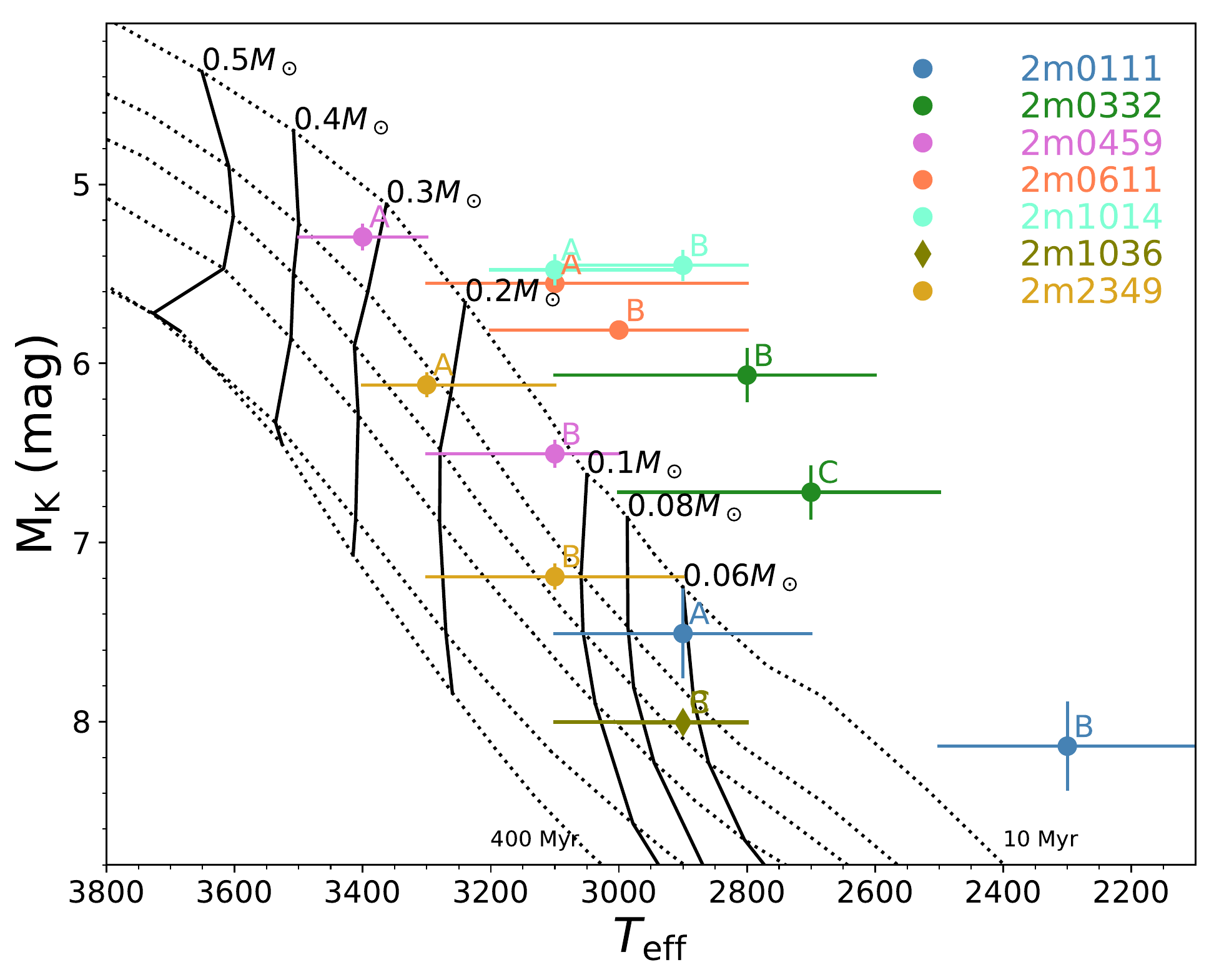}
\caption{Isochrones from the BHAC \citep{baraffe_new_2015} evolutionary models plotted as 2MASS $K$-band magnitudes as a function of effective temperature. The isochrones span ages 10, 20, 30, 50, 120 and 400 Myrs, going from youngest at the top to oldest at the bottom of the figure. Mass-tracks at $0.06, 0.08, 0.1, 0.2, 0.3, 0.4$ and $0.5\,M_{\odot}$ are displayed going across the isochrones. The binary pairs are included in the figure as coloured circles where the absolute magnitude is calculated from the resolved $K'$-band magnitudes in Table~\ref{tab:phot} with either Gaia DR2 parallaxes or statistical distances from \cite{malo_bayesian_2013}. Effective temperatures are listed in Table~\ref{tab:spt}, where the {\it Top panel} show the effective temperatures derived from the H$_2$O--K2-indices, and the {\it Bottom panel} show the effective temperature estimated from the $K$-band continuum fitted to theoretical models. The bottom panel indicates younger ages, and that the uncertainty for the effective temperature derived from the H$_2$O--K2-index is underestimated. The J1036 system is depicted by diamonds in the plots, and is the only binary pair for which a dynamical mass-estimate exists, estimated to be $0.24\pm 0.07\,M_{\odot}$ for each individual component \citep{calissendorff_discrepancy_2017}.
}
\label{fig:HRD}
\end{figure}

Theoretical evolutionary models can also provide additional information on the age and mass of a system when compared to dynamical masses. One of our binaries in our sample has its dynamical mass constrained from an orbital fit already, J1036 \citep{calissendorff_discrepancy_2017}, with each binary component having an individual mass of $0.24\pm 0.07 \,M_{\odot}$, which we can use to directly compare to the measured luminosities, temperatures and masses in a Hertzsprung-Russel diagram (HRD). We compare the $K$-band magnitudes and effective temperatures of the individual binary components for J1036 to six different isochrones of ages 10, 20, 30, 50, 120 and 400 Myrs from the BHAC15 models \citep{baraffe_new_2015} shown in Figure~\ref{fig:HRD}. The isochrone ages are chosen to better represent the different YMG ages associated with the systems of interest, and that after $\sim 400$ Myr the objects enter the main sequence track, and the models do not change much at these low stellar masses. Several mass-tracks are displayed in the figure, highlighting theoretical masses at 0.5, 0.4, 0.3, 0.2, 0.1, 0.08 and $0.06\,M_{\odot}$. \cite{calissendorff_discrepancy_2017} found a discrepancy of $\approx 30\,\%$ between the dynamical mass-estimate and the theoretical masses from the evolutionary models, with the models under-predicting the mass. We find similar results for the B component when we compare the $K$-band magnitudes to measured effective temperature from the H$_2$O--K2-index measurements, that the models suggest much lower mass compared to the dynamical mass estimate. We also find from the isochrone fit a younger age for the system than anticipated. However, for the C component the isochronal fit is consistent with the dynamical mass estimate. It is unlikely that the temperature is so different for the two components in the J1036 system, as they are of equal mass and brightness. We attribute the observed discrepancy in our H2O-K2 index measurements to issues with the astrometric calculations and brightness scaling for the B component, as its halo is not entirely within the field of view, and that the uncertainty is underestimated.

Nevertheless, when comparing the effective temperature of $2800 - 3100$ K obtained from the $K$-band continuum spectral fit instead, we find the components to be more consistent between each other, and within $2-\sigma$ of the dynamical mass-estimate and age-range of the Ursa Majoris moving group or field population. The spectral EW feature analysis indicates for the system to have lower surface gravity compared to the older field population, but also on the highest compared to the other binaries in our sample, which is consistent with the age-estimate of young field or $\approx 400$ Myr Ursa Majoris moving group.

The absolute magnitude in $K$-band is here calculated using the Gaia DR2 parallax for the J1036 system, corresponding to a distance of $d = 19.75 \pm 0.06$ pc, and slightly less than the distance used in \cite{calissendorff_discrepancy_2017}, albeit having a negligible impact on the absolute magnitude ($\leq 0.04\,\Delta$mag) and isochrone fit.

We also include the other binary pairs in the HRD in Figure~\ref{fig:HRD}, but without dynamical mass-estimates the results become more difficult to gauge. Nevertheless, we find most objects to be consistent with young ages from the isochrone analysis, but that most of them have under-predicted masses when we apply the effective temperatures derived from the H$_2$O--K2-index.

\section{Summary \& Conclusions}\label{sec:summary}
We provide a near-infrared spectral analysis of seven close visual M-dwarf binaries from the AstraLux multiplicity survey, constraining spectral types for all resolved components in our study. Although the binary components are spatially resolved with the help of adaptive optics, PSF overlaps typically persist. We use PSF matching techniques to minimise cross-contamination of the spectra resulting from this overlap. Nonetheless, imperfections in the PSF matching can still potentially cause some residual contamination in the closest or highest-contrast binaries. We further probe the targets for low surface gravity by scrutinising EW measurements of gravity sensitive alkali lines and other signs of youth, finding them all to have some or several indicators for being of young age. The validation of young ages for these systems will be particularly beneficial for dynamical mass-studies in the future, as these systems may be used as exemplary benchmarks for calibrating theoretical evolutionary models with.

Kinematics including proper motions and trigonometric parallax measurements from Gaia DR2 are missing for the J0111, J0459 and J1014 binaries. Regardless, kinematic measurements from the ground and spectroscopic distance measurements place these three systems as very high probable members of YMGs \citep{malo_bayesian_2013}. Our EW analysis are consistent with this prediction that they are indeed young, but future Gaia data releases may provide firmer constraints on not only the kinematics of these systems but also YMGs and associations as a whole.

The main purpose of this study was to characterise symbolic binaries which can be used as benchmarks to test against theoretical models, which tend to systematically under predict masses for low-mass stars $M \leq 0.5\,M_{\odot}$ \citep{hillenbrand_assessment_2004, calissendorff_discrepancy_2017, rodet_dynamical_2018}. We find our sample to be consistent with young ages, ranging $\sim 20 - 400$ Myrs according to the YMG membership probabilities obtained from the BANYAN $\Sigma$-tool. Compared to the other AstraLux sample of seven low-mass M-dwarf binaries from \cite{bergfors_characterisation_2016}, our sample is overall younger. For the \cite{bergfors_characterisation_2016} sample, we find with the BANYAN $\Sigma$-tool that only the J06134539-2352077 binary system belongs to the Argus moving group, while the rest are either part of the field or have insufficient data for a rigorous assessment of Bayesian YMG membership probabilities. Thus, the systems studied in our sample will become part of a larger sample of well-studied young low-mass M-dwarf binaries, which together with the \cite{bergfors_characterisation_2016} extends 14 characterised binaries over a large age-span that may be tested against evolutionary models once orbital constraints can be placed. As the AstraLux is an ongoing astrometric monitoring survey, and since these systems have predicted orbital periods of just a few years to a few decades, we can expect to soon be able to constrain the dynamical masses for some of these emblematic binaries.


\begin{acknowledgements} 
We thank the referee, B\'{a}rbara Rojas-Ayala, for the insightful comments and advice that helped improve the manuscript. M.J. gratefully acknowledges funding from the Knut and Alice Wallenberg foundation. This publication makes use of data products from the Two Micron All Sky Survey, which is a joint project of the University of Massachusetts and the Infrared Processing and Analysis Center/California Institute of Technology, funded by the National Aeronautics and Space Administration and the National Science Foundation. This research has benefitted from the Montreal Brown Dwarf and Exoplanet Spectral Library, maintained by Jonathan Gagn\'{e}. This work has made use of data from the European Space Agency (ESA) mission {\it Gaia} (\url{https://www.cosmos.esa.int/gaia}), processed by the {\it Gaia} Data Processing and Analysis Consortium (DPAC, \url{https://www.cosmos.esa.int/web/gaia/dpac/consortium}). Funding for the DPAC has been provided by national institutions, in particular the institutions participating in the {\it Gaia} Multilateral Agreement.

\end{acknowledgements}

\bibliographystyle{aa} 
\bibliography{references-Mdwarf_binaries.bib}      

\begin{thebibliography}{62}
\expandafter\ifx\csname natexlab\endcsname\relax\def\natexlab#1{#1}\fi

\bibitem[{Allers {et~al.}(2007)Allers, Jaffe, Luhman, Liu, Wilson, Skrutskie,
  Nelson, Peterson, Smith, \& Cushing}]{allers_characterizing_2007}
Allers, K.~N., Jaffe, D.~T., Luhman, K.~L., {et~al.} 2007, ApJ, 657, 511

\bibitem[{Baraffe {et~al.}(2015)Baraffe, Homeier, Allard, \&
  Chabrier}]{baraffe_new_2015}
Baraffe, I., Homeier, D., Allard, F., \& Chabrier, G. 2015, A\&A, 577, A42

\bibitem[{Bergfors {et~al.}(2016)Bergfors, Brandner, Bonnefoy, Schlieder,
  Janson, Henning, \& Chauvin}]{bergfors_characterisation_2016}
Bergfors, C., Brandner, W., Bonnefoy, M., {et~al.} 2016, Mon. Not. R. Astron.
  Soc., 456, 2576, arXiv: 1511.09119

\bibitem[{Bergfors {et~al.}(2010)Bergfors, Brandner, Janson, Daemgen, Geissler,
  Henning, Hippler, Hormuth, Joergens, \& Köhler}]{bergfors_lucky_2010}
Bergfors, C., Brandner, W., Janson, M., {et~al.} 2010, A\&A, 520, A54, arXiv:
  1006.2377

\bibitem[{Bonnefoy {et~al.}(2009)Bonnefoy, Chauvin, Dumas, Lagrange, Beust,
  Desort, Teixeira, Ducourant, Beuzit, \& Song}]{bonnefoy_young_2009}
Bonnefoy, M., Chauvin, G., Dumas, C., {et~al.} 2009, A\&A, 506, 799

\bibitem[{Bonnefoy {et~al.}(2014)Bonnefoy, Chauvin, Lagrange, Rojo, Allard,
  Pinte, Dumas, \& Homeier}]{bonnefoy_library_2014}
Bonnefoy, M., Chauvin, G., Lagrange, A.-M., {et~al.} 2014, A\&A, 562, A127,
  arXiv: 1306.3709

\bibitem[{Bonnet {et~al.}(2003)Bonnet, Strobele, Biancat-Marchet, Brynnel,
  Conzelmann, Delabre, Donaldson, Farinato, Fedrigo, Hubin, Kasper, \&
  Kissler-Patig}]{bonnet_implementation_2003}
Bonnet, H., Strobele, S., Biancat-Marchet, F., {et~al.} 2003, Waikoloa,
  Hawai'i, United States, 329

\bibitem[{Bouy {et~al.}(2008)Bouy, Martín, Brandner, Forveille, Delfosse,
  Huélamo, Basri, Girard, Zapatero~Osorio, Stumpf, Ghez, Valdivielso, Marchis,
  Burgasser, \& Cruz}]{bouy_follow-up_2008}
Bouy, H., Martín, E.~L., Brandner, W., {et~al.} 2008, A\&A, 481, 757

\bibitem[{Bowler {et~al.}(2015)Bowler, Shkolnik, Liu, Schlieder, Mann, Dupuy,
  Hinkley, Crepp, Johnson, Howard, Flagg, Weinberger, Aller, Allers, Best,
  Kotson, Montet, Herczeg, Baranec, Riddle, Law, Nielsen, Wahhaj, Biller, \&
  Hayward}]{bowler_planets_2015}
Bowler, B.~P., Shkolnik, E.~L., Liu, M.~C., {et~al.} 2015, ApJ, 806, 62

\bibitem[{Burgasser {et~al.}(2006)Burgasser, Reid, Siegler, Close, Allen,
  Lowrance, \& Gizis}]{burgasser_not_2006}
Burgasser, A.~J., Reid, I.~N., Siegler, N., {et~al.} 2006,
  arXiv:astro-ph/0602122, arXiv: astro-ph/0602122

\bibitem[{Calissendorff \& Janson(2018)}]{calissendorff_improving_2018}
Calissendorff, P. \& Janson, M. 2018, A\&A, 615, A149

\bibitem[{Calissendorff {et~al.}(2019)Calissendorff, Janson, Asensio-Torres, \&
  Köhler}]{calissendorff_spectral_2019}
Calissendorff, P., Janson, M., Asensio-Torres, R., \& Köhler, R. 2019, A\&A,
  627, A167

\bibitem[{Calissendorff {et~al.}(2017)Calissendorff, Janson, Köhler, Durkan,
  Hippler, Dai, Brandner, Schlieder, \&
  Henning}]{calissendorff_discrepancy_2017}
Calissendorff, P., Janson, M., Köhler, R., {et~al.} 2017, A\&A, 604, A82

\bibitem[{Close {et~al.}(2005)Close, Lenzen, Guirado, Nielsen, Mamajek,
  Brandner, Hartung, Lidman, \& Biller}]{close_dynamical_2005}
Close, L.~M., Lenzen, R., Guirado, J.~C., {et~al.} 2005, Nature, 433, 286

\bibitem[{Collaboration(2016)}]{gaia_collaboration_gaia_2016}
Collaboration, G. 2016, A\&A, 595, A1, arXiv: 1609.04153

\bibitem[{Collaboration {et~al.}(2018)Collaboration, Brown, Vallenari, Prusti,
  de~Bruijne, Babusiaux, \& Bailer-Jones}]{gaia_collaboration_gaia_2018}
Collaboration, G., Brown, A. G.~A., Vallenari, A., {et~al.} 2018, A\&A, 616,
  A1, arXiv: 1804.09365

\bibitem[{Covey {et~al.}(2010)Covey, Lada, Román-Zúñiga, Muench, Forbrich,
  \& Ascenso}]{covey_age_2010}
Covey, K.~R., Lada, C.~J., Román-Zúñiga, C., {et~al.} 2010, ApJ, 722, 971

\bibitem[{Cushing {et~al.}(2008)Cushing, Marley, Saumon, Kelly, Vacca, Rayner,
  Freedman, Lodders, \& Roellig}]{cushing_atmospheric_2008}
Cushing, M.~C., Marley, M.~S., Saumon, D., {et~al.} 2008, ApJ, 678, 1372

\bibitem[{Cushing {et~al.}(2005)Cushing, Rayner, \&
  Vacca}]{cushing_infrared_2005}
Cushing, M.~C., Rayner, J.~T., \& Vacca, W.~D. 2005, ApJ, 623, 1115

\bibitem[{Daemgen {et~al.}(2007)Daemgen, Siegler, Reid, \&
  Close}]{daemgen_discovery_2007}
Daemgen, S., Siegler, N., Reid, I.~N., \& Close, L.~M. 2007, ApJ, 654, 558

\bibitem[{David {et~al.}(2019)David, Hillenbrand, Gillen, Cody, Howell,
  Isaacson, \& Livingston}]{david_age_2019}
David, T.~J., Hillenbrand, L.~A., Gillen, E., {et~al.} 2019, ApJ, 872, 161,
  arXiv: 1901.05532

\bibitem[{Dressing \& Charbonneau(2013)}]{dressing_occurrence_2013}
Dressing, C.~D. \& Charbonneau, D. 2013, ApJ, 767, 95

\bibitem[{Duchêne \& Kraus(2013)}]{duchene_stellar_2013}
Duchêne, G. \& Kraus, A. 2013, Annu. Rev. Astron. Astrophys., 51, 269

\bibitem[{Dupuy {et~al.}(2010)Dupuy, Liu, Bowler, Cushing, Helling, Witte, \&
  Hauschildt}]{dupuy_studying_2010}
Dupuy, T.~J., Liu, M.~C., Bowler, B.~P., {et~al.} 2010, ApJ, 721, 1725

\bibitem[{Eisenhauer {et~al.}(2003)Eisenhauer, Abuter, Bickert,
  Biancat-Marchet, Bonnet, Brynnel, Conzelmann, Delabre, Donaldson, Farinato,
  Fedrigo, Genzel, Hubin, Iserlohe, Kasper, Kissler-Patig, Monnet, Roehrle,
  Schreiber, Stroebele, Tecza, Thatte, \& Weisz}]{eisenhauer_sinfoni_2003}
Eisenhauer, F., Abuter, R., Bickert, K., {et~al.} 2003, Waikoloa, Hawai'i,
  United States, 1548

\bibitem[{Freudling {et~al.}(2013)Freudling, Romaniello, Bramich, Ballester,
  Forchi, García-Dabló, Moehler, \& Neeser}]{freudling_automated_2013}
Freudling, W., Romaniello, M., Bramich, D.~M., {et~al.} 2013, A\&A, 559, A96

\bibitem[{Gagné \& Faherty(2018)}]{gagne_banyan_2018}
Gagné, J. \& Faherty, J.~K. 2018, ApJ, 862, 138, arXiv: 1805.11715

\bibitem[{Gagné {et~al.}(2015)Gagné, Faherty, Cruz, Lafreniére, Doyon, Malo,
  Burgasser, Naud, Artigau, Bouchard, Gizis, \& Albert}]{gagne_banyan_2015}
Gagné, J., Faherty, J.~K., Cruz, K.~L., {et~al.} 2015, ApJS, 219, 33

\bibitem[{Gagné {et~al.}(2014{\natexlab{a}})Gagné, Lafrenière, Doyon, Malo,
  \& Artigau}]{gagne_banyan_2014}
Gagné, J., Lafrenière, D., Doyon, R., Malo, L., \& Artigau, E.
  2014{\natexlab{a}}, ApJ, 783, 121

\bibitem[{Gagné {et~al.}(2014{\natexlab{b}})Gagné, Lafrenière, Doyon, Malo,
  \& Artigau}]{gagne_banyan_2014-1}
Gagné, J., Lafrenière, D., Doyon, R., Malo, L., \& Artigau, E.
  2014{\natexlab{b}}, ApJ, 798, 73

\bibitem[{Gaidos {et~al.}(2014)Gaidos, Mann, Lépine, Buccino, James, Ansdell,
  Petrucci, Mauas, \& Hilton}]{gaidos_trumpeting_2014}
Gaidos, E., Mann, A.~W., Lépine, S., {et~al.} 2014, Monthly Notices of the
  Royal Astronomical Society, 443, 2561

\bibitem[{Gillon {et~al.}(2016)Gillon, Jehin, Lederer, Delrez, de~Wit,
  Burdanov, Van~Grootel, Burgasser, Triaud, Opitom, Demory, Sahu,
  Bardalez~Gagliuffi, Magain, \& Queloz}]{gillon_temperate_2016}
Gillon, M., Jehin, E., Lederer, S.~M., {et~al.} 2016, Nature, 533, 221

\bibitem[{Goodwin {et~al.}(2008)Goodwin, Nutter, Kroupa, Ward-Thompson, \&
  Whitworth}]{goodwin_relationship_2008}
Goodwin, S.~P., Nutter, D., Kroupa, P., Ward-Thompson, D., \& Whitworth, A.~P.
  2008, A\&A, 477, 823

\bibitem[{Gorlova {et~al.}(2003)Gorlova, Meyer, Rieke, \&
  Liebert}]{gorlova_gravity_2003}
Gorlova, N.~I., Meyer, M.~R., Rieke, G.~H., \& Liebert, J. 2003, ApJ, 593, 1074

\bibitem[{Hillenbrand \& White(2004)}]{hillenbrand_assessment_2004}
Hillenbrand, L.~A. \& White, R.~J. 2004, ApJ, 604, 741

\bibitem[{Janson {et~al.}(2014{\natexlab{a}})Janson, Bergfors, Brandner,
  Bonnefoy, Schlieder, Kohler, Hormuth, Henning, \&
  Hippler}]{janson_orbital_2014}
Janson, M., Bergfors, C., Brandner, W., {et~al.} 2014{\natexlab{a}}, ApJS, 214,
  17, arXiv: 1409.1550

\bibitem[{Janson {et~al.}(2014{\natexlab{b}})Janson, Bergfors, Brandner,
  Kudryavtseva, Hormuth, Hippler, \& Henning}]{janson_astralux_2014}
Janson, M., Bergfors, C., Brandner, W., {et~al.} 2014{\natexlab{b}}, ApJ, 789,
  102, arXiv: 1406.0535

\bibitem[{Janson {et~al.}(2007)Janson, Brandner, Lenzen, Close, Nielsen,
  Hartung, Henning, \& Bouy}]{janson_improved_2007}
Janson, M., Brandner, W., Lenzen, R., {et~al.} 2007, A\&A, 462, 615

\bibitem[{Janson {et~al.}(2012)Janson, Hormuth, Bergfors, Brandner, Hippler,
  Daemgen, Kudryavtseva, Schmalzl, Schnupp, \& Henning}]{janson_astralux_2012}
Janson, M., Hormuth, F., Bergfors, C., {et~al.} 2012, ApJ, 754, 44, arXiv:
  1205.4718

\bibitem[{Johnson {et~al.}(2012)Johnson, Gazak, Apps, Muirhead, Crepp,
  Crossfield, {Tabetha Boyajian}, von Braun, Rojas-Ayala, Howard, Covey,
  Schlawin, Hamren, Morton, Marcy, \& Lloyd}]{johnson_characterizing_2012}
Johnson, J.~A., Gazak, J.~Z., Apps, K., {et~al.} 2012, The Astronomical
  Journal, 143, 111

\bibitem[{Jones {et~al.}(2015)Jones, White, Boyajian, Schaefer, Baines,
  Ireland, Patience, Brummelaar, McAlister, Ridgway, Sturmann, Sturmann,
  Turner, Farrington, \& Goldfinger}]{jones_ages_2015}
Jones, J., White, R.~J., Boyajian, T., {et~al.} 2015, ApJ, 813, 58

\bibitem[{Kirkpatrick {et~al.}(2006)Kirkpatrick, Barman, Burgasser, McGovern,
  McLean, Tinney, \& Lowrance}]{kirkpatrick_discovery_2006}
Kirkpatrick, J.~D., Barman, T.~S., Burgasser, A.~J., {et~al.} 2006, ApJ, 639,
  1120

\bibitem[{Klutsch {et~al.}(2014)Klutsch, Freire~Ferrero, Guillout, Frasca,
  Marilli, \& Montes}]{klutsch_reliable_2014}
Klutsch, A., Freire~Ferrero, R., Guillout, P., {et~al.} 2014, A\&A, 567, A52

\bibitem[{Kraus \& Hillenbrand(2007)}]{kraus_stellar_2007}
Kraus, A.~L. \& Hillenbrand, L.~A. 2007, AJ, 134, 2340

\bibitem[{Köhler {et~al.}(2012)Köhler, Ratzka, \&
  Leinert}]{kohler_orbits_2012}
Köhler, R., Ratzka, T., \& Leinert, C. 2012, A\&A, 541, A29, arXiv: 1203.6270

\bibitem[{Malo {et~al.}(2013)Malo, Doyon, Lafrenière, Artigau, Gagné, Baron,
  \& Riedel}]{malo_bayesian_2013}
Malo, L., Doyon, R., Lafrenière, D., {et~al.} 2013, ApJ, 762, 88

\bibitem[{Mamajek {et~al.}(2010)Mamajek, Kenworthy, Hinz, \&
  Meyer}]{mamajek_discovery_2010}
Mamajek, E.~E., Kenworthy, M.~A., Hinz, P.~M., \& Meyer, M.~R. 2010, The
  Astronomical Journal, 139, 919

\bibitem[{Manara {et~al.}(2013)Manara, Testi, Rigliaco, Alcalá, Natta,
  Stelzer, Biazzo, Covino, Covino, Cupani, D’Elia, \&
  Randich}]{manara_x-shooter_2013}
Manara, C.~F., Testi, L., Rigliaco, E., {et~al.} 2013, A\&A, 551, A107

\bibitem[{Mentuch {et~al.}(2008)Mentuch, Brandeker, van Kerkwijk, Jayawardhana,
  \& Hauschildt}]{mentuch_lithium_2008}
Mentuch, E., Brandeker, A., van Kerkwijk, M.~H., Jayawardhana, R., \&
  Hauschildt, P.~H. 2008, Astrophysical Journal, 689, 1127, arXiv: 0808.3584

\bibitem[{Mizuki {et~al.}(2018)Mizuki, Kuzuhara, Mede, Schlieder, Janson,
  Brandt, Hirano, Narita, Wisniewski, Yamada, Biller, Bonnefoy, Carson,
  McElwain, Matsuo, Turner, Mayama, Akiyama, Uyama, Nakagawa, Kudo, Kusakabe,
  Hashimoto, Abe, Brander, Egner, Feldt, Goto, Grady, Guyon, Hayano, Hayashi,
  Hayashi, Henning, Hodapp, Ishii, Iye, Kandori, Knapp, Kwon, Miyama, Morino,
  Moro-Martin, Nishimura, Pyo, Serabyn, Suenaga, Suto, Suzuki, Takahashi,
  Takami, Takato, Terada, Thalmann, Watanabe, Takami, Usuda, \&
  Tamura}]{mizuki_orbital_2018}
Mizuki, T., Kuzuhara, M., Mede, K., {et~al.} 2018, ApJ, 865, 152

\bibitem[{Montet {et~al.}(2015)Montet, Bowler, Shkolnik, Deck, Wang, Horch,
  Liu, Hillenbrand, Kraus, \& Charbonneau}]{montet_dynamical_2015}
Montet, B.~T., Bowler, B.~P., Shkolnik, E.~L., {et~al.} 2015, ApJ, 813, L11

\bibitem[{Muirhead {et~al.}(2014)Muirhead, Becker, Feiden, Rojas-Ayala,
  Vanderburg, Price, Thorp, Law, Riddle, Baranec, Hamren, Schlawin, Covey,
  Johnson, \& Lloyd}]{muirhead_characterizing_2014}
Muirhead, P.~S., Becker, J., Feiden, G.~A., {et~al.} 2014, ApJS, 213, 5

\bibitem[{Muirhead {et~al.}(2012)Muirhead, Hamren, Schlawin, Rojas-Ayala,
  Covey, \& Lloyd}]{muirhead_characterizing_2012}
Muirhead, P.~S., Hamren, K., Schlawin, E., {et~al.} 2012, ApJ, 750, L37

\bibitem[{Rayner {et~al.}(2009)Rayner, Cushing, \&
  Vacca}]{rayner_infrared_2009}
Rayner, J.~T., Cushing, M.~C., \& Vacca, W.~D. 2009, ApJS, 185, 289

\bibitem[{Riaz {et~al.}(2006)Riaz, Gizis, \& Harvin}]{riaz_identification_2006}
Riaz, B., Gizis, J.~E., \& Harvin, J. 2006, AJ, 132, 866, arXiv:
  astro-ph/0606617

\bibitem[{Rodet {et~al.}(2018)Rodet, Bonnefoy, Durkan, Beust, Lagrange,
  Schlieder, Janson, Grandjean, Chauvin, Messina, Maire, Brandner, Girard,
  Delorme, Biller, Bergfors, Lacour, Feldt, Henning, Boccaletti, Bouquin,
  Berger, Monin, Udry, Peretti, Segransan, Allard, Homeier, Vigan, Langlois,
  Hagelberg, Menard, Bazzon, Beuzit, Delboulbe, Desidera, Gratton, Lannier,
  Ligi, Maurel, Mesa, Meyer, Pavlov, Ramos, Rigal, Roelfsema, Salter, Samland,
  Schmidt, Stadler, \& Weber}]{rodet_dynamical_2018}
Rodet, L., Bonnefoy, M., Durkan, S., {et~al.} 2018, A\&A, 618, A23, arXiv:
  1806.05491

\bibitem[{Rojas-Ayala {et~al.}(2012)Rojas-Ayala, Covey, Muirhead, \&
  Lloyd}]{rojas-ayala_metallicity_2012}
Rojas-Ayala, B., Covey, K.~R., Muirhead, P.~S., \& Lloyd, J.~P. 2012, ApJ, 748,
  93, arXiv: 1112.4567

\bibitem[{Schlieder {et~al.}(2014)Schlieder, Bonnefoy, Herbst, Lépine, Berger,
  Henning, Skemer, Chauvin, Rice, Biller, Girard, Lagrange, Hinz, Defrère,
  Bergfors, Brandner, Lacour, Skrutskie, \&
  Leisenring}]{schlieder_characterization_2014}
Schlieder, J.~E., Bonnefoy, M., Herbst, T.~M., {et~al.} 2014, ApJ, 783, 27

\bibitem[{Sembach \& Savage(1992)}]{sembach_observations_1992}
Sembach, K.~R. \& Savage, B.~D. 1992, ApJS, 83, 147

\bibitem[{Skrutskie {et~al.}(2006)Skrutskie, Cutri, Stiening, Weinberg,
  Schneider, Carpenter, Beichman, Capps, Chester, Elias, Huchra, Liebert,
  Lonsdale, Monet, Price, Seitzer, Jarrett, Kirkpatrick, Gizis, Howard, Evans,
  Fowler, Fullmer, Hurt, Light, Kopan, Marsh, McCallon, Tam, Van~Dyk, \&
  Wheelock}]{skrutskie_two_2006}
Skrutskie, M.~F., Cutri, R.~M., Stiening, R., {et~al.} 2006, AJ, 131, 1163

\bibitem[{Torres {et~al.}(2008)Torres, Quast, Melo, \&
  Sterzik}]{torres_young_2008}
Torres, C. A.~O., Quast, G.~R., Melo, C. H.~F., \& Sterzik, M.~F. 2008,
  arXiv:0808.3362 [astro-ph], arXiv: 0808.3362

\bibitem[{Zuckerman \& Song(2004)}]{zuckerman_young_2004}
Zuckerman, B. \& Song, I. 2004, Annu. Rev. Astron. Astrophys., 42, 685

\end{thebibliography}
\end{document}